\documentstyle[aps,pre,prabib,epsf,floats,amssymb,multicol]{revtex}

\newcommand{\bi}{\begin{itemize}\setlength{\itemsep}{0pt}}
\newcommand{\ei}{\end{itemize}}
\newcommand{\be}{\begin{equation}}
\newcommand{\ee}{\end{equation}}
\newcommand{\bea}{\begin{eqnarray}}
\newcommand{\eea}{\end{eqnarray}}
\def\(#1){(\ref{#1})}

\newcommand{\bfig}[3]{\begin{figure}\vspace*{#2}\begin{center}%
\leavevmode\epsfxsize #1\epsfbox{#3.eps}}
\newcommand{\efig}[2]{\end{center}\caption{#2%
\label{fig:#1}}\end{figure}}

\newcommand{\E}{E}
\newcommand{\Emax}{E_{\rm max}}
\newcommand{\ommin}{\om_{\rm min}}

\newcommand{\Gn}{\Gamma_0}
\newcommand{\Gloc}{\Gamma_{\rm th}}
\newcommand{\G}{\Gamma}
\newcommand{\lc}{{l_{\rm y}}}
\newcommand{\rh}{\rho}
\newcommand{\gam}{\gamma}
\newcommand{\gamc}{{\gamma_{\rm c}}}
\newcommand{\gamdot}{{\dot{\gamma}}}
\newcommand{\eq}{_{\rm eq}}

\newcommand{\g}{_{\rm g}}
\newcommand{\sig}{\sigma}
\newcommand{\sigy}{\sigma_{\rm y}}

\newcommand{\ZZ}[2]{Z(#1,#2)}
\newcommand{\visc}{\eta}

\newcommand{\om}{\omega}
\newcommand{\Gcomp}{G^*}
\newcommand{\dynvisc}{{\eta^*}}
\newcommand{\bkz}{_{\rm BKZ}}
\newcommand{\Gzero}{G_0}
\newcommand{\Grho}{G_\rho}
\newcommand{\de}{\Delta E}

\newcommand{\half}{{1\over 2}}
\newcommand{\deriv}[1]{{\partial\over\partial{#1}}}
\newcommand{\lav}{\left\langle}
\newcommand{\rav}{\right\rangle}

\newcommand{\ltappr}{\lesssim}
\newcommand{\gtappr}{\gtrsim}

\newcommand{\ie}{{\it i.e.}}
\newcommand{\eg}{{\it e.g.}}
\newcommand{\lhs}{lhs}
\newcommand{\rhs}{rhs}

\draft
\begin{document}

\title{Rheological constitutive equation for model of soft glassy materials}

\author{Peter Sollich}

\address{Department of Physics, University of Edinburgh, Edinburgh EH9
3JZ, U.K., {\tt P.Sollich@ed.ac.uk}}

\maketitle

\begin{abstract}
We solve exactly and describe in detail a simplified scalar model for
the low frequency shear rheology of foams, emulsions, slurries,
etc. [P.\ Sollich, F.\ Lequeux, P.\ H{\'{e}}braud, M.E.\ Cates, Phys.\
Rev.\ Lett. {\bf 78}, 2020 (1997)].  The model attributes similarities
in the rheology of such ``soft glassy materials'' to the shared
features of structural disorder and metastability. By focusing on the
dynamics of mesoscopic elements, it retains a generic character.
Interactions are represented by a mean-field noise temperature $x$,
with a glass transition occurring at $x=1$ (in appropriate units).
The exact solution of the model takes the form of a constitutive
equation relating stress to strain history, from which all rheological
properties can be derived. For the linear response, we find that both
the storage modulus $G'$ and the loss modulus $G''$ vary with
frequency as $\omega^{x-1}$ for $1<x<2$, becoming flat near the glass
transition. In the glass phase, aging of the moduli is predicted. The
steady shear flow curves show power law fluid behavior for $x<2$, with
a nonzero yield stress in the glass phase; the Cox-Merz rule does not
hold in this non-Newtonian regime. Single and double step strains
further probe the nonlinear behavior of the model, which is not well
represented by the BKZ relation. Finally, we consider measurements of
$G'$ and $G''$ at finite strain amplitude $\gam$. Near the glass
transition, $G''$ exhibits a maximum as $\gam$ is increased in a
strain sweep. Its value can be strongly overestimated due to nonlinear
effects, which can be present even when the stress response is very
nearly harmonic. The largest strain $\gamc$ at which measurements
still probe the linear response is predicted to be roughly
frequency-independent.
\end{abstract}

\pacs{PACS numbers: 83.20.-d, 83.70.Hq, 05.40+j. To appear in
{\em Physical Review E} (July 1998).}

\begin{multicols}{2}

\section{Introduction}

Many soft materials, such as foams, emulsions, pastes and slurries,
have intriguing rheological properties. Experimentally, there is a
well-developed phenomenology for such systems: their nonlinear flow
behavior is often fit to the form $\sigma = A + B \dot \gamma^n$ where
$\sigma$ is shear stress and $\dot\gamma$ strain rate.  This is the
Herschel-Bulkeley equation~\cite{Holdsworth93,Dickinson92}; or (for
$A=0$) the ``power-law
fluid''~\cite{Holdsworth93,Dickinson92,BarHutWal89}.  For the same
materials, linear or quasi-linear viscoelastic measurements often
reveal storage and loss moduli $G'(\omega)$, $G''(\omega)$ in nearly
constant ratio ($G''/G'$ is usually about 0.1) with a frequency
dependence that is either a weak power law (clay slurries, paints,
microgels) or negligible (tomato paste, dense emulsions, dense
multilayer vesicles, colloidal glasses)
\cite{MacMarSmeZha94,KetPruGra88,KhaSchneArm88,%
MasBibWei95,PanRouVuiLuCat96,HofRau93,MasWei95}.  This behavior
persists down to the lowest accessible frequencies (about $10^{-3}$--1
Hz depending on the system), in apparent contradiction to linear
response theory, which requires that $G''(\om)$ should be an odd
function of $\om$.  This behavior could in principle be due to slow
relaxation modes below the experimentally accessible frequency range
(see Fig.~\ref{fig:sample_spectrum}).  Each of those would cause a
drop in $G'(\omega)$ and a bump in $G''(\omega)$ as the
frequency is tracked downward. However, where the search for system
specific candidates for such slow modes has been carried out (for
the case of foams and dense emulsions, for example, 
see~\cite{BuzLuCat95}), it has not
yielded viable candidates; it therefore seems worthwhile to look for
more generic explanations of the observed behavior.

Indeed, the fact that similar anomalous rheology should be seen in
such a wide range of soft materials suggests a common cause. In
particular, the frequency dependence indicated above points strongly
to the generic presence of slow ``glassy'' dynamics persisting to
arbitrarily small frequencies. This feature is found in several other
contexts~\cite{Bouchaud92,MonBou96,BouDea95}, such as the dynamics of
elastic manifolds in random media~\cite{VinMarChe96,LeDouVin95}.  The
latter is suggestive of rheology: charge density waves, vortices,
contact lines, etc.\ can ``flow'' in response to an imposed
``stress''.

In a previous letter~\cite{SolLeqHebCat97} it was argued that glassy
dynamics is a natural consequence of two properties shared by all the
soft materials mentioned above: {\em structural disorder} and {\em
metastability}. In such ``soft glassy materials'' (SGMs), thermal
motion alone is not enough to achieve complete structural
relaxation. The system has to cross energy barriers (for example those
associated with rearrangement of droplets in an emulsion) that are
very large compared to typical thermal energies. It therefore adopts a
disordered, metastable configuration even when (as in a monodisperse
emulsion or foam) the state of least free energy would be
ordered~\cite{emulsion_inherent_metastability}.  The importance of
structural disorder has previously been noted in more specific
contexts~\cite{MasBibWei95,BuzLuCat95,WeaFor94,%
LacGreLevMasWei96,OkuKaw95,Durian95,Durian97}, but its unifying role
in rheological modeling can be more easily appreciated by focusing on
the class of SGMs as a whole.

In Ref.~\cite{SolLeqHebCat97}, a minimal, scalar model for the generic
rheology of SGMs was introduced, which incorporates the above
ideas. We refer to this model as the ``soft glassy rheology'' (SGR)
model in the following. The main contribution of the present
publication is the exact solution of this model; at the same time, we
also provide more detailed analytical and numerical support for the
results announced in~\cite{SolLeqHebCat97}. The exact solution is in
the form of a
\end{multicols}
\newpage
\twocolumn
\noindent
constitutive equation relating the (shear) stress at a given time to
the strain history. We use this to study a range of linear and
nonlinear rheological properties of the model; qualitative comparisons
with experimental data show that these capture many generic
rheological characteristics of SGMs. We do not attempt more
quantitative fits to experimental data for specific materials because
the model in its present form is almost certainly too over-simplified
for this purpose. We do however hope to carry out such a more
quantitative study in future work, once the remaining ambiguities in
the interpretation of the model parameters (see
Sec.~\ref{sec:interpretation}) have been clarified and some of the
improvements suggested in Sec.~\ref{sec:conclusion} have been
incorporated into the model.

We introduce the SGR model in Sec.~\ref{sec:model}, along with
Bouchaud's glass model on which it builds. Sec.~\ref{sec:ce} contains
our main result, the constitutive
equation. Its predictions in the linear response regime are discussed
in Sec.~\ref{sec:linear_response}, while in Sec.~\ref{sec:nonlinear}
we analyse several nonlinear scenarios including steady shear flow,
shear startup, large step strains and large oscillatory strains.  The
physical significance and interpretation of the various parameters of
the SGR model is not obvious; in Sec.~\ref{sec:interpretation} we
discuss in more detail the ``noise temperature'' $x$ and ``attempt
frequency'' $\Gn$ of the model. Our results are summarized in
Sec.~\ref{sec:conclusion}.
\bfig{9cm}{0mm}{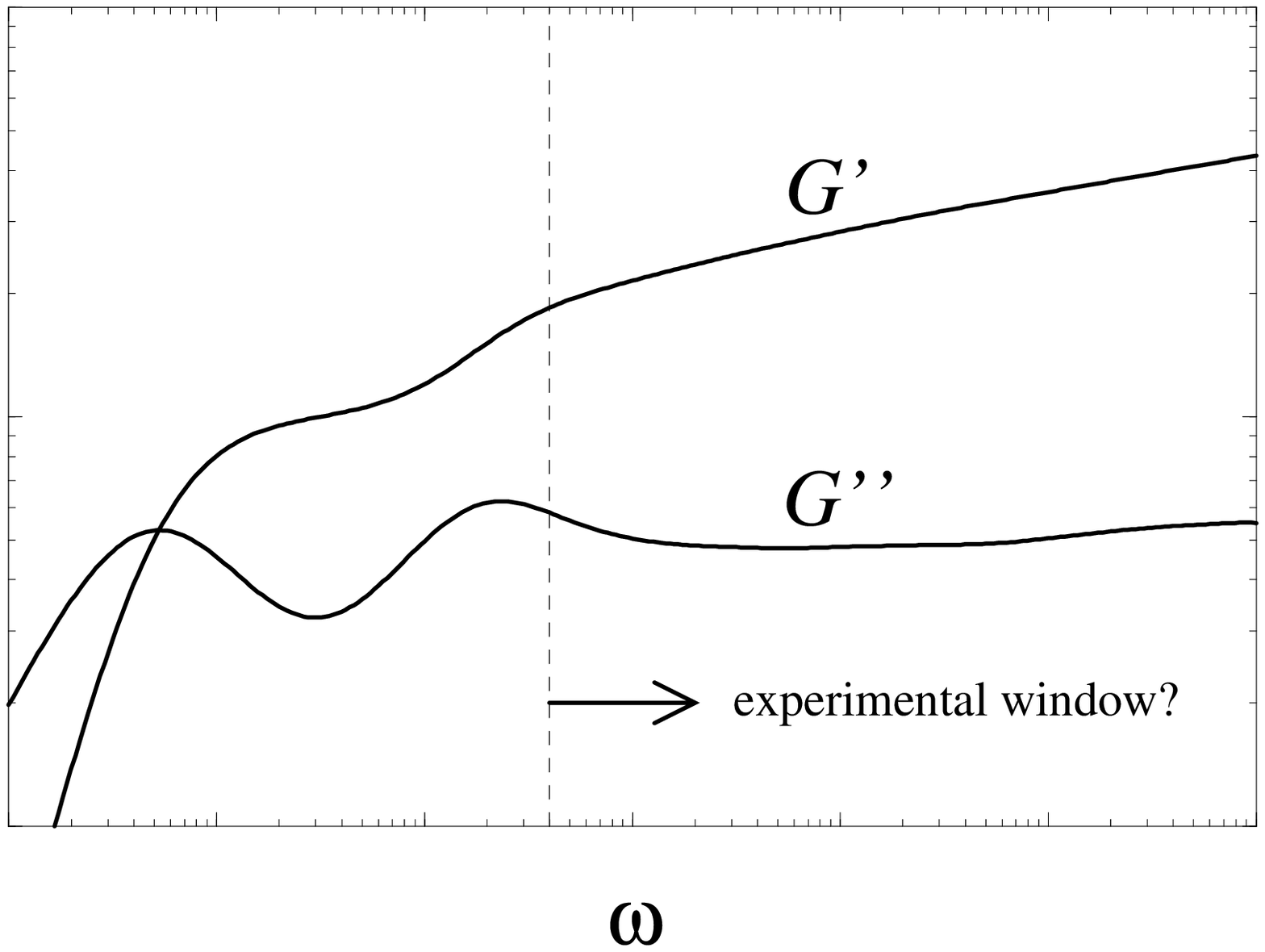}
\efig{sample_spectrum}{Sketch of frequency dependence of linear
moduli, showing possible slow relaxation modes at frequencies below
the measurement window.}

\section{The SGR model}
\label{sec:model}

The SGR model is a phenomenological model that aims to explain the
main features of SGM rheology (both linear and nonlinear) as described
above. To apply to a broad range of materials, such a model needs to
be reasonably generic.  It should therefore incorporate only a minimal
number of features common to all SGMs, leaving aside as much system
specific detail as possible. One important feature is the
``glassiness'', \ie, the effects of structural disorder and
metastability. We model this using a fairly intuitive picture of a
glass: it consists of local ``elements'' (we will be more specific later
about what we mean by these in the context of SGMs) which are trapped
in ``cages'' formed by their neighbors so that they cannot
move. Occasionally, however, a rearrangement of the elements may be
possible, due to thermal activation, for example. Glass models of this
kind are commonly referred to as ``trap models'' and have been studied
by a large number of authors (see \eg\
Refs.~\cite{MonBou96,BouComMon95,OdaMatHiw94,%
ArkBaes94,RicBaes90,Vilgis90,Baessler87,HauKeh87}).  An alternative to
such models would be, for example, mode-coupling
theories~\cite{Goetze91,GoetSjoe92} which, at least in their simplest
form, neglect all (thermally) activated processes. 
We prefer trap models for our purposes,	because they are simpler and
also generally more physically transparent~\cite{strong_fragile_note}.

\subsection{Bouchaud's glass model}
\label{subsec:bouchaud}

Bouchaud formalized the above intuitive trap picture of a glass into a
one-element model~\cite{Bouchaud92,MonBou96}: an individual
element ``sees'' an energy landscape of traps of various depths $E$;
when activated, it can ``hop'' to another trap. Bouchaud assumed that
such hopping processes are due to thermal fluctuations. In SGMs,
however, this is unlikely as $k_{\rm B}T$ is very small compared to
typical trap depths $E$ (see Sec.~\ref{sec:interpretation}). The SGR
model assumes instead that the ``activation'' in SGMs is due to {\em
interactions}: a rearrangement somewhere in the material can propagate
and cause rearrangements elsewhere. In a mean-field spirit, this
coupling between elements is represented by an {\em effective
temperature} (or noise level) $x$. This idea is fundamental to the SGR
model.

The equation of motion for the probability of finding
an element in a trap of depth $E$ at time $t$
is~\cite{Bouchaud92,MonBou96,Epositive_footnote}
\be
\deriv{t}P(E,t) =  -\Gn e^{-E/x}\,P(E,t) + \G(t)\,\rh(E)
\label{bouchaud_eq_motion}
\ee
In the first term on the \rhs, which describes elements hopping out of
their current traps, $\Gn$ is an attempt frequency for hops, and
$\exp(-E/x)$ is the corresponding activation factor. The second term
represents the state of these elements directly after a hop. Bouchaud
made the simplest possible assumption that the depth of the new trap
is completely independent of that of the old one; it is simply
randomly chosen from some ``prior'' distribution of trap depths
$\rho(E)$. The rate of hopping into traps of depth $E$ is then
$\rho(E)$ times the overall hopping rate, given by
\be
\G(t)=\Gn \lav e^{-E/x}\rav_P = \Gn \int dE\  P(E,t)\, e^{-E/x}
\label{hop_rate}
\ee
Bouchaud's main insight was that the model~\(bouchaud_eq_motion) can
describe a glass transition {\em if the density of deep traps has an
exponential tail}, $\rho(E)\sim\exp(-E/x\g)$, say. Why is this?
The steady state of eq.~\(bouchaud_eq_motion), if one exists, is given
by $P\eq(E)\propto\exp(E/x)\rho(E)$; the Boltzmann factor $\exp(E/x)$
(no minus here because trap depths are measured from zero {\em
downwards}) is proportional to the average time spent in a trap of
depth $E$. At $x=x\g$, it just cancels the exponential decay of
$\rho(E)$, and so the supposed equilibrium distribution $P\eq(E)$
tends to a constant for large $E$; it is not normalizable.  This means
that, for $x\leq x\g$, the system does not have a steady state; it is
(``weakly'') non-ergodic and ``ages'' by evolving into deeper and deeper
traps~\cite{Bouchaud92,MonBou96}. The model~\(bouchaud_eq_motion)
therefore has a {\em glass transition} at $x=x\g$.

With Bouchaud's model, we have a good candidate for describing in a
relatively simple way the glassy features of SGMs. Its disadvantages
for our purposes are: (i) The assumption of an exponentially decaying
$\rho(E)$ is rather arbitrary in our context.  It can be justified in
systems with ``quenched'' (\ie, fixed) disorder, such as spin glasses,
using extreme value statistics (see \eg~\cite{BouMez97}), but it is
not obvious how to extend this argument to SGMs. (ii) The exponential
form of the activation factor in~\(bouchaud_eq_motion) was chosen by
analogy with thermal activation. But for us, $x$ describes effective
noise arising from interactions, so this analogy is by no means
automatic, and functional forms other than exponential could also be
plausible. In essence, we view (i) {\em together with} (ii) as a
phenomenological way of describing a system with a glass transition.

\subsection{Incorporating deformation and flow}
\label{subsec:model}

To describe deformation and flow, the SGR model~\cite{SolLeqHebCat97}
incorporates strain degrees of freedom into Bouchaud's glass model. A
generic SGM is conceptually subdivided into a large number of {\em
mesoscopic regions}, and these form the ``elements'' of the model. By
mesoscopic we mean that these regions must be (i) small enough for a
macroscopic piece of material to contain a large number of them,
allowing us to describe its behavior as an {\em average} over
elements; and (ii) large enough so that deformations on the scale of
an element can be described by an elastic strain variable. For a
single droplet in a foam, for example, this would not be possible
because of its highly non-affine deformation; in this case, the
element size should therefore be at least a few droplet diameters.
The size of the elements is chosen as the unit length to avoid
cumbersome factors of element volume in the expressions below.
We emphasize that the subdivision into mesoscopic elements is merely a
conceptual tool for obtaining a suitably coarse-grained description of
a SGM. The elements should not be thought of as sharply-defined
physical entities, but rather as somewhat diffuse ``blobs'' of
material. Their size simply represents a coarse-graining length scale
whose order of magnitude is fixed by the two requirements (i) and (ii)
above.

We denote by $l$ the local shear strain of an element (more generally,
the deformation would have to be described by a tensor, but we choose
a simple scalar description).  To see how $l$ evolves as the system is
sheared, consider first the behavior of a foam or dense emulsion. The
droplets in an element will initially deform elastically from the local
equilibrium configuration, giving rise to a stored elastic energy (due
to surface tension, in this example~\cite{WeaFor94}).  This continues
up to a yield point, characterized by a strain $\lc$, whereupon the
droplets rearrange to new positions in which they are less deformed,
thus relaxing stress.  The mesoscopic strain $l$ {\em measured from
the nearest equilibrium position} (\ie, the one the element would
relax to if there were no external stresses) is then again
zero. As the macroscopic strain $\gam$ is increased, $l$ therefore
executes a ``saw-tooth'' kind of motion~\cite{sawtooth_note}.
Neglecting nonlinearities before yielding, the local shear stress is
given by $kl$, with $k$ an elastic constant; the yield point
defines a maximal elastic energy $\E=\half k\lc^2$.  The effects of
structural disorder are modeled by assuming a {\em distribution} of
such yield energies $\E$, rather than a single value common to all
elements.  A similar description obviously extends to many others of
the soft materials mentioned above.

To make the connection to Bouchaud's glass model, yield events can
be viewed as ``hops'' out of a trap (or potential well), and the yield
energy $E$ is thereby identified with the trap depth. As before, we
assume that yields (hops) are activated by interactions between
different elements, resulting in an effective temperature $x$. The
activation barrier is now $E-\half kl^2$, the difference between the
typical yield energy and the elastic energy already stored in the
element.

For the behavior of elements in between rearrangements, the simplest
assumption is that their strain changes along with the macroscopically
imposed strain $\gam$. This means that, yield events apart, the {\em
shear rate} is homogeneous throughout the material; spatial
fluctuations of the shear rate are neglected in what can be viewed as
a further mean-field approximation. The SGR model therefore applies
only to materials which can support macroscopically homogeneous flows
(at least in the range of shear rates of practical interest). In fact,
we regard this requirement as a working definition of what is meant by
a ``soft'' glassy material. A ``hard'' glassy material, on the other
hand, might fail by fracture and strong strain localization rather
than by homogeneous flow. Whether a link exists between this
distinction and the classification of structural glasses into fragile
versus strong~\cite{strong_fragile_note} is not clear to us at present.

While the SGR model assumes a spatially homogeneous {\em strain rate},
it does admit inhomogeneities in the local {\em strain} $l$ and {\em
stress} $\sig=kl$~\cite{elastic_manifolds_note}. These arise because
different elements generally yield at different times. To describe the
state of the system at a given time, we therefore now need to know the
joint probability of finding an element with a yield energy $E$ {\em
and} a local strain $l$. Within the SGR model~\cite{SolLeqHebCat97},
this probability evolves in time according to
\bea
\lefteqn{\deriv{t}P(E,l,t) = }\nonumber\\
& & - \gamdot\deriv{l}P 
- \Gn\, e^{-(E-\half kl^2)/x}\,P + \G(t)\,\rh(E)\delta(l)
\label{basic}
\eea
The first term on the \rhs\ describes the motion of the elements
between rearrangements, with a local strain rate equal to the
macroscopic one, $\dot{l}=\gamdot$. The interaction-activated yielding
of elements (which is assumed to be an instantaneous process on the
timescales of interest to us) is reflected in the second term.  The
last term incorporates two assumptions about the properties of an
element just after yielding: It is unstrained ($l=0$) and has a new
yield energy $E$ randomly chosen from $\rho(E)$, \ie, uncorrelated
with its previous one.  Finally, the total yielding rate is given by
the analog of~\(hop_rate),
\bea
\G(t) &=& \Gn \lav e^{-(E-\half kl^2)/x} \rav_P \nonumber\\
      &=& \Gn \int dE \,dl\ P(E,l,t) \, e^{-(E-\half kl^2)/x} 
\eea
Eq.~\(basic) tells us how the state of the system, described by
$P(E,l,t)$, evolves for a given imposed macroscopic strain
$\gam(t)$. What we mainly care about is of course the rheological
response, \ie, the macroscopic stress. This is given by the average of
the local stresses
\be
\sig(t)=k\lav l\rav_P \equiv k \int\! dE \,dl\ P(E,l,t)\,l
\label{stress_def}
\ee

Eqs.~(\ref{basic}-\ref{stress_def}) define the SGR model, a minimal
model for the rheology of SGMs: It incorporates both the ``glassy''
features arising from structural disorder (captured in the
distribution of yield energies $E$ and local strains $l$) and the
``softness'': for large macroscopic strains, the material flows
because eventually all elements yield.  An intuitive picture of the
dynamics of the SGR model can be obtained by viewing each element as a
``particle'' moving in a one-dimensional piecewise quadratic
potential, with noise-induced hops which become increasingly likely
near the edge of a potential well (see Fig.~\ref{fig:wells}). This
also shows the hysteresis effects associated with yielding: Once a hop
to a new well has taken place, a finite strain reversal is in general
needed before a particle will hop back to its old
well~\cite{bistable_footnote}.
\bfig{8cm}{0cm}{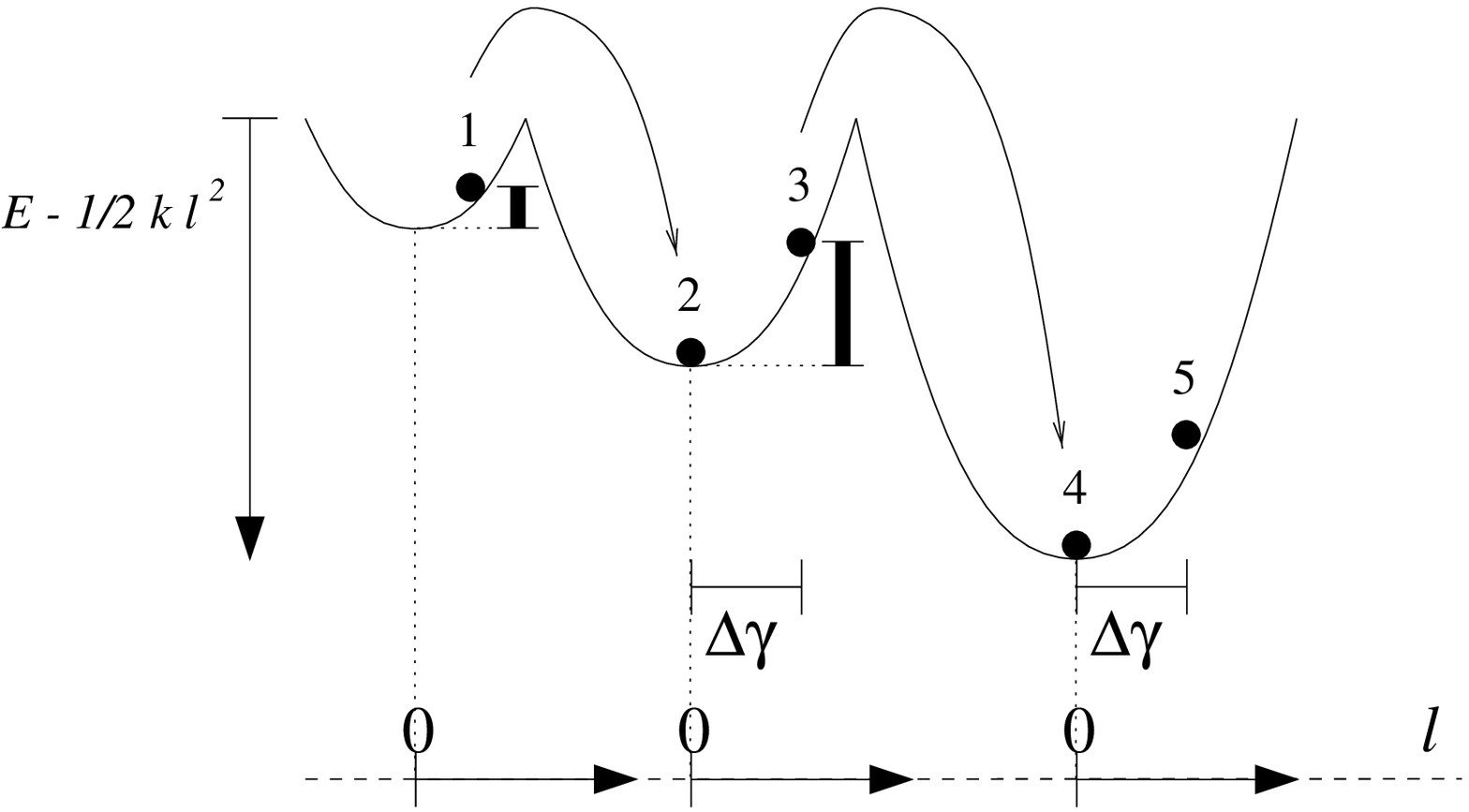}
\efig{wells}{Potential well picture of the dynamics of the SGR
model. Note that the relative horizontal displacement of the quadratic
potential wells is arbitrary; each has its own independent zero for the
scale of the local strain $l$. The solid vertical bars indicate the
energy dissipated in the ``hops'' (yielding events) from 1 to 2 and 3
to 4, respectively.}

Before moving on to the exact solution of the SGR model, we briefly
mention some of its limitations.  Among the most serious of these is
the assumption that the noise temperature $x$ and the attempt
frequency $\Gn$ are constant parameters of the model. In general, they
may be expected to depend on the imposed shear rate $\gamdot$, for
example, or in fact have their own intrinsic time evolution.  In
particular, it must be born in mind when interpreting our results
below that the effective noise temperature $x$ is not a parameter that
we can easily tune from the outside; rather, we expect it to be
determined self-consistently by the interactions in the system.  We
discuss these points in some detail in Sec.~\ref{sec:interpretation},
where we also speculate on the physical origin of the model parameters
$x$ and $\Gn$. Within the SGR model, the ``prior'' density of yield
energies, $\rho(E)$, is likewise taken to be a constant. This implies the
assumption that the structure of the material considered is not
drastically altered by an imposed flow, and excludes effects such as
shear-induced crystallization.

The SGR model is also essentially a low-frequency model.
This is due to our assumption that each element behaves purely
elastically until it yields and a rearrangement takes place.  In
reality, the rheological response of an element will be more
complex. After the application of a strain, for example, there may be
a fast relaxation of the local stress from its instantaneous value,
due to local relaxation processes. In a foam, for example, these might
correspond to small shifts of the bubble positions; in the language of
mode-coupling theory, they could be described as
$\beta$-relaxations~\cite{GoetSjoe92,beta_relax_footnote}. Such local
stress relaxation processes are expected to take place much faster
than actual yielding events, which involve a more drastic
reorganization of the structure of the material. For frequencies
smaller than the attempt frequency for yielding, $\omega\ltappr\Gn$,
they can therefore be neglected.  This then implies that the
elastic properties that we ascribe to local elements are those that
apply once all fast local stress relaxation processes are complete.
We have also neglected viscous contributions to the local stress; in
foams, for example, these are due to the flow of water and surfactant
caused by the deformation of the elements. In the low frequency regime
of interest to us, such viscous effects are again insignificant (see
\eg~\cite{BuzLuCat95}), whereas at high frequencies the
model~(\ref{basic}-\ref{stress_def}) would have to be modified
appropriately to yield sensible predictions.

Another restriction of the model is the assumption that the elastic
constant $k$ is the same for all elements.  This may not be
appropriate, for example, for strongly polydisperse materials; we plan
to investigate the effects of variable $k$ in future work. We have
also made the simplifying assumption that an element is always
unstrained directly after yielding. Interaction between neighboring
elements may however frustrate the relaxation to the new equilibrium
state; we discuss briefly in Sec.~\ref{subsec:linear_frustration} how
this feature can be incorporated into the model.

Finally, the treatment of energy dissipation during yield events
within the SGR model may 
also have to be refined. This can be seen by expressing the work done
on the system in the following way: We multiply the equation of
motion~\(basic) by the elastic energy $\half kl^2$ of an element and
integrate over $l$ and $\E$. Integration by parts of the $\gamdot$
term then just gives the stress~\(stress_def), hence
\be 
\sig\gamdot = \frac{d}{dt} \half\lav k\l^2\rav + \Gn \half\lav kl^2
e^{-(E-\half kl^2)/x}\rav
\label{energy_balance}
\ee 
where the averages are over $P(\E, l, t)$. The \lhs\ is the rate of
energy input into the system. The first term on the \rhs, which is a
complete time differential, describes the part of this energy that is
stored as elastic energy of the elements. The second term, which is
always non-negative, is the dissipative part.  It is just the average
over all elements of their yielding rate times the energy dissipated
in a rearrangement, which we read off as $\half kl^2$. This means
that within the model, every rearrangement dissipates exactly the
elastic energy stored within the element when it yields (see
Fig.~\ref{fig:wells}). 

In general, this is not implausible. But it implies that some
rearrangements---those of unstrained ($l=0$) elements---have no
dissipation associated with them~\cite{no_diss_in_eq}.  In reality,
however, the local reorganization of a material during {\em any} yield
event would always be expected to dissipate {\em some} energy. How
much might depend, for example, on the height of the activation
barrier for yielding, $E-\half kl^2$. The model in its present form
does not capture such effects; in fact, the yield energies $E$ do not
feature in the energy balance~\(energy_balance) except through their
effect on the yielding rates. This exposes a related limitation of the
model: On physical grounds, one would expect that elements with a
larger yield energy $\E$ may have a more stable configuration with
lower total energy (for example, an arrangement of droplets in an
emulsion with a lower total surface energy). The average value of $\E$
(which increases during aging, for
example~\cite{Bouchaud92,MonBou96}), should then also occur in the
energy balance~\(energy_balance). This is not accounted for in the
model in its present form.

\section{Constitutive equation}
\label{sec:ce}

To simplify the following analysis of the model, we choose appropriate
units for energy and time; a convenient choice is such that
$x\g=\Gn=1$.  From the definition of the glass transition temperature,
this implies that the density of yield energies has the form
$\rh(\E)=\exp[-\E(1+f(\E))]$ with $f(\E)\to 0$ for $\E\to\infty$. For
our numerical investigations below we use the simplest $\rh(E)$ of
this form, which is purely exponential
\be
\rh(E)=\exp(-E)
\label{exp_rho}
\ee
Analytical results, on the other hand, hold for general $\rh(E)$
unless otherwise stated. We eliminate a final parameter from the model
by setting $k=1$; this can always be achieved by a rescaling of the
stress $\sig$ and the strain variables $\gamma$ and $l$.  With this
choice of units, it becomes clear that the SGR model is in fact rather
parsimonious: apart from scale factors, its predictions are determined
by a single parameter, the effective noise temperature
$x$~\cite{rho_E_footnote}.

Note that in our chosen units, typical yield strains $\sqrt{2E/k}$ are
of order one. Experimentally, SGMs generally have yield stresses of at
most a few percent (see
\eg~\cite{MasWei95,HebLeqMunPin97,MasBibWei96}); the necessary
rescaling of our results for strain variables should be born in mind
when comparing to experimental data.  For example, a strain rate
$\gamdot=1$ in our units corresponds to $\gamdot=\overline{\lc} \Gn$
in dimensional units, with $\overline{\lc}=(x\g/k)^{1/2}$ a typical
(``a priori'', \ie, sampled from $\rho(E)$) yield strain. For a
specific material, the three scale parameters $x\g$, $k$ and $\Gn$ of
the SGR model could be estimated from measurements of a yield strain,
a shear modulus and a viscosity, for example.

The derivation of the exact constitutive equation (CE) for the SGR model is
given App.~\ref{app:ce}. For simplicity, we impose the mild
restriction that the initial state is completely unstrained, \ie,
$\gam(t=0)=0$ and
\be
P(E,l,t=0)=P_0(E)\delta(l)
\label{unstrained_initial}
\ee
Our central result then relates the stress at time $t$ to the strain
history $\gam(t')$ ($0<t'<t$) by the CE
\bea
\sig(t)&=&\gam(t)\,\Gzero(\ZZ{t}{0}) \nonumber\\
& & + 
\int_0^t dt'\ \G(t')\,[\gam(t)-\gam(t')]\,\Grho(\ZZ{t}{t'})
\label{ce}
\eea
with the yielding rate $\G(t)$ determined from
\be
1=\Gzero(\ZZ{t}{0}) + \int_0^t dt' \ \G(t')\, \Grho(\ZZ{t}{t'})
\label{ce_Gam}
\ee
Here the functions
\bea
\Gzero(z) &=& \int dE\, P_0(E) \exp\left(-ze^{-E/x}\right) \nonumber\\
\Grho(z) &=& \int dE\, \rh(E) \exp\left(-ze^{-E/x}\right)
\label{Gzero_rho_def}
\eea
describe the purely noise induced decay of the stress.  This decay is
however governed not simply by the time interval between a change in
macroscopic strain at $t'$ and a stress measurement at $t$, but by an
``effective time interval'' $z=\ZZ{t}{t'}$ given by
\be
\ZZ{t}{t'} = \int_{t'}^t dt''
\exp\left\{\left[\gam(t'')-\gam(t')\right]^2/2x\right\}
\label{Z_def}
\ee
One reads off that $\ZZ{t}{t'}\geq t-t'$; the effective time interval
is always {\em greater} than the actual time interval, and the more so
the larger the changes in strain $\gam(t'')$ from its value at the
earlier time $t'$.  This implies a faster decay of the stress, and so
$\ZZ{t}{t'}$ can be said to describe strain-induced yielding (in other
words, shear-thinning). In fact, a look at~(\ref{ce},\ref{ce_Gam})
confirms that {\em all} nonlinear effects within the model arise from
this dependence of the effective time interval $\ZZ{t}{t'}$ on the
macroscopic strain history $\gam(t'')$.

The CE~(\ref{ce},\ref{ce_Gam}) can be most easily understood by
viewing the yielding of elements as a birth-death process: Each time
an element yields, it ``dies'' and is ``reborn'' with $l=0$. In
between such events, its local strain just follows the changes in
global strain $\gam(t)$. If an element was last reborn at time $t'$,
its local strain at time $t$ is therefore $l=\gam(t)-\gam(t')$.  Since
we set $k=1$, this is also its contribution to the stress.  The first
term on the \rhs\ of~(\ref{ce},\ref{ce_Gam}) is the contribution of
elements which have ``survived'' from time $0$ to $t$; they do so with
the ``survival probability'' $\Gzero(\ZZ{t}{0})$.  The second term
collects the contribution from all elements which have yielded at
least once between time $0$ and $t$, and were last reborn at $t'$. The
number of such elements is proportional to the rate of ``rebirths'' at
$t'$, \ie, the yielding rate $\G(t')$, and the corresponding survival
probability $\Grho(\ZZ{t}{t'})$. Note that there are two different
survival probabilities here, given by $\Gzero$ and $\Grho$,
respectively. The difference arises from the fact that these
probabilities are in fact averages over the distribution of yield
energies, as expressed by~\(Gzero_rho_def).  For elements that have
survived from $t'=0$, this distribution is $P_0(E)$, while for
elements that have yielded at least once, it is
$\rh(E)$.

The glassy features of the SGR model as discussed in
Sec.~\ref{subsec:bouchaud} are reflected
in the CE~(\ref{ce},\ref{ce_Gam}), in particular in
the asymptotic behavior of $\Grho(z)$. For the simple exponential
form~\(exp_rho) of $\rho(E)$, one easily finds that
$\Grho(z)=x!\,z^{-x}$ asymptotically. 
As shown in Appendix~\ref{app:Grho_asympt}, the same 
behavior holds for general $\rho(E)$, in the sense that
\bea
\lim_{z\to\infty} \ \Grho(z)\,z^{x+\epsilon} & = & \infty \nonumber\\
\lim_{z\to\infty} \ \Grho(z)\,z^{x-\epsilon} & = & 0
\label{Grho_asympt}
\eea
for any arbitrarily small $\epsilon>0$.  We shall refer to this
property by saying that $\Grho(z)$ decays asymptotically as $z^{-x}$
up to ``sub-power law factors''. Unless otherwise specified, all power
laws referred to in the following hold for general $\rho(E)$, up to
such sub-power law factors.

Consider now the case where strain-induced yielding can be neglected,
such that $\ZZ{t}{t'}=t-t'$. This is always true for sufficiently
small strain amplitudes.  Below the glass transition ($x<1$), the time
integral $\int_0^t dt'\, \Grho(t-t')$ of the response function
$\Grho(\ZZ{t}{t'})=\Grho(t-t')$ in~\(ce) then diverges in the limit
$t\to\infty$. Compatible with the intuitive notion of a glass phase,
this means that the system has a very long memory (of the kind that
has been described as ``weak long term
memory''~\cite{CugKur93,CugKur95}) and is (weakly~\cite{Bouchaud92})
non-ergodic.  This can lead to rather
intricate aging behavior, which we plan to explore in future work. For
the purpose of the present paper---with the exception of a brief
discussion in Sec.~\ref{subsec:linear_glass}---we focus on situations
where the system is ergodic. These include the regime above the glass
transition, $x>1$, and the case of steady shear flow  for all
noise temperatures $x$ (strain-induced yielding here 
restores ergodicity even for $x<1$).
In the former case, a choice needs to be made
for the initial distribution of yield energies. We consider the
simplest case where this is the equilibrium distribution at the given
$x$,
\be
P_0(E) = P\eq(E) = \G\eq \exp(E/x)\rho(E)
\label{eq_initial}
\ee
Correspondingly, we write $G_0(z)=G\eq(z)$.  The function
$\Grho(z)$ is then related to the derivative of $G\eq(z)$ by
\be
\Grho(z)=-\G\eq^{-1}G\eq'(z)
\label{Grho_Geq}
\ee
with a proportionality constant given by the equilibrium yielding rate
\be
\G\eq^{-1}=\int \! dE \ \rho(E)\,\exp(E/x)=\int_0^\infty \! dz \ \Grho(z)
\label{Gam_eq}
\ee

\section{Linear response}
\label{sec:linear_response}

\subsection{Above the glass transition}

The simplest characterization of the rheological behavior of the SGR
model is through its linear rheology. This describes the stress
response to small shear strain perturbations around the equilibrium
state. As such, it is well defined (\ie, time-independent) a priori
only above the glass transition, $x>1$ (see however
Sec.~\ref{subsec:linear_glass}).

To linear order in the applied strain $\gam(t)$, the effective time
interval $\ZZ{t}{t'}=t-t'$. In the linear regime, all yield events
are therefore purely noise-induced rather than strain-induced.
Correspondingly, the yielding rate as determined from~\(ce_Gam) is
simply $\G(t)=\G\eq$, as can be confirmed from
eqs.~(\ref{Grho_Geq},\ref{Gam_eq}).  The expression~\(ce) for the
stress can then be simplified to the familiar form
\be
\sig(t)=\int_0^t\! dt'\,\gamdot(t') \, G\eq(t-t')
\label{lin_response}
\ee
As expected for an equilibrium situation, the response is
time-translation invariant~\cite{lower_limit_zero_note}, with
$G\eq(t)$ being the linear stress response to a unit step strain at
$t=0$.  The dynamic modulus is obtained by Fourier transform,
\be
\Gcomp(\om) = i\omega \int_0^\infty dt\  e^{-i\omega t}\, G\eq(t) =
\lav\frac{i\om\tau}{i\om\tau+1}\rav\eq
\label{dyn_modulus}
\ee
This an average over Maxwell modes with relaxation times $\tau$. For
an element with yield energy $E$, $\tau=\exp(E/x)$ is just its average
lifetime, \ie, the average time between rearrangements. The relaxation
time spectrum therefore follows from the equilibrium distribution of
energies, $P\eq(E)\propto\exp(E/x)\rho(E)$. Because of the exponential
tail of $\rho(E)$, it has a power-law tail $P\eq(\tau)\sim
\tau^{-x}$ (for $\tau\gg 1$, up to sub-power law factors).  As $x$
decreases towards the glass transition, this long-time part of the
spectrum becomes increasingly dominant and causes anomalous low
frequency behavior of the moduli, as shown in
Fig.~\ref{fig:linear_moduli}:
\bea
G'  \sim \omega^2 \    & \mbox{\ for $3<x$}, & \quad
    \sim \omega^{x-1}\   \mbox{\ for $1<x<3$} \nonumber\\
G'' \sim \omega\ \; \, & \mbox{\ for $2<x$}, & \quad
    \sim \omega^{x-1}\   \mbox{\ for $1<x<2$}
\label{G_low_freq}
\eea
\bfig{8.5cm}{0cm}{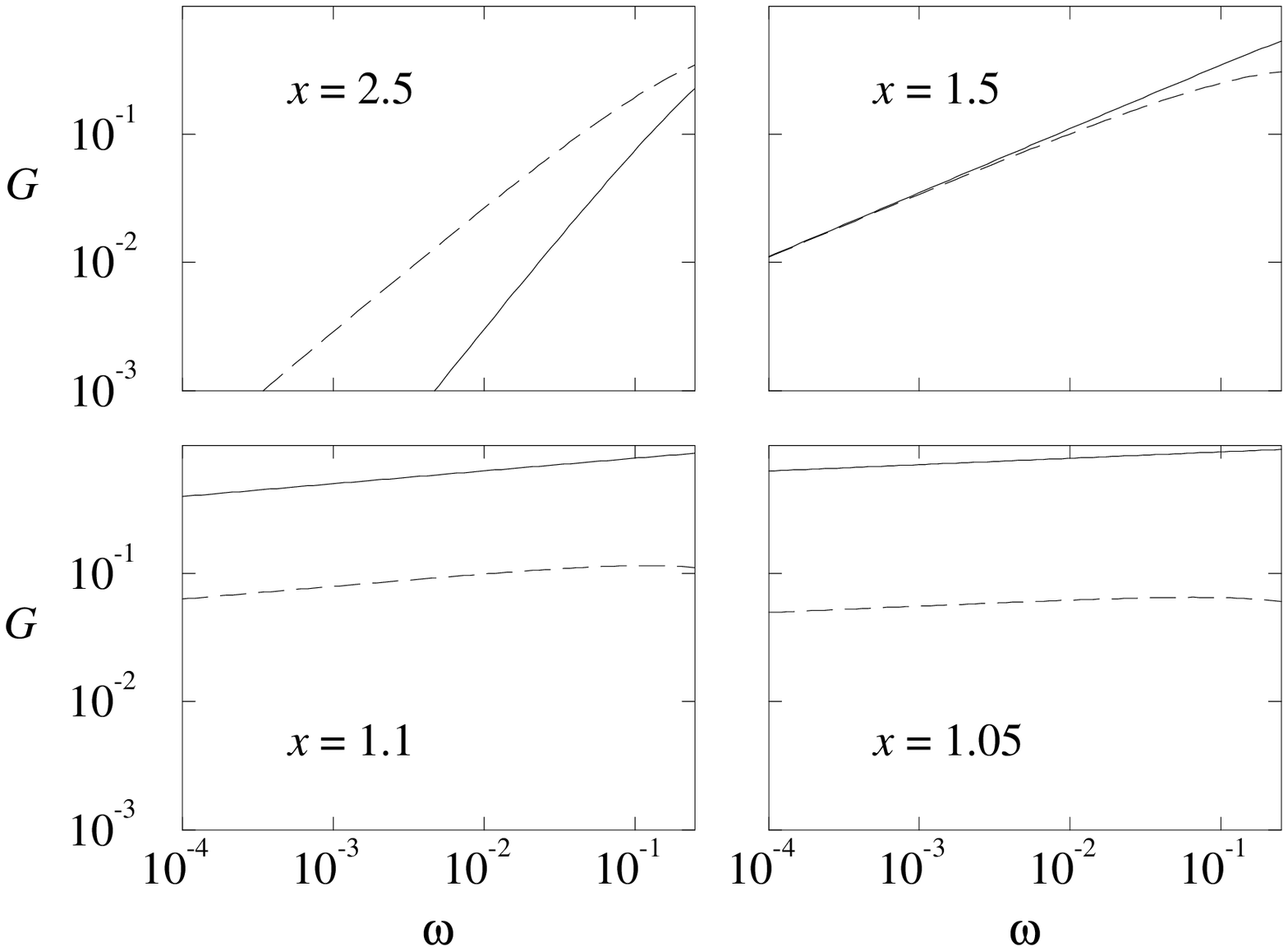} 
\efig{linear_moduli}{Linear moduli $G'$ (solid line) and $G''$
(dashed) vs frequency $\om$ at various noise temperatures $x$. We only
show the behavior in the low frequency regime $\om\ltappr 1$, where
the predictions of the SGR model are expected to be physically
relevant. The high frequency behavior (predicted as
$G'\approx \mbox{const}$, $G''\sim\om^{-1}$) is not
realistic because the model neglects local viscous effects (among
others) which
can become important in this regime.}
For $x>3$ the system is Maxwell-like at low frequencies, whereas for
$2<x<3$ there is an anomalous power law in the elastic modulus.  Most
interesting is the regime $1<x<2$, where $G'$ and $G''$ have constant
ratio; both vary as $\omega^{x-1}$.  Behavior like this is observed in
a number of soft materials
\cite{MacMarSmeZha94,KetPruGra88,KhaSchneArm88,MasBibWei95,MasWei95}.
Moreover, the frequency exponent approaches zero as $x\to 1$,
resulting in essentially constant values of $G''$ and $G'$, as
reported in dense emulsions, foams, and onion
phases~\cite{KhaSchneArm88,MasBibWei95,PanRouVuiLuCat96}. Note,
however, that the ratio $G''/G'\sim x-1$ becomes small as the glass
transition is approached. This increasing dominance of the elastic
response $G'$ prefigures the onset of a yield stress for $x<1$
(discussed below). It does not mean, however, that the loss modulus
$G''$ for fixed (small) $\om$ always decreases with $x$; in fact, it
first {\em increases} strongly as $x$ is lowered and only starts
decreasing close to the glass transition (when $x-1\sim
|\ln\om|^{-1}$). The reason for this crossover is that the relaxation
time $\tau(\lav E\rav\eq)=\exp(\lav\E\rav\eq/x)$ corresponding to the mean
equilibrium energy $\lav\E\rav\eq\sim (x-1)^{-1}$ eventually becomes
greater than $\om^{-1}$.

\subsection{Glass phase}
\label{subsec:linear_glass}

The above linear results only apply above the glass transition
($x>1$), where there is a well defined equilibrium state around which
small perturbations can be made. However, if a cutoff $\Emax$ on the
yield energies is introduced (which is physically reasonable because
yield strains cannot be arbitrarily large), an equilibrium state also
exists for $x<1$, \ie, below the glass transition.  (Strictly
speaking, with the cutoff imposed there is no longer a true glass
phase; but if the energy cutoff is large enough, its qualitative
features are expected to be still present.) One then finds for the low
frequency behavior of the linear moduli:
\be
G' \approx\mbox{const.} \qquad G''\sim \om^{x-1}
\label{G_glass}
\ee
\bfig{9cm}{0cm}{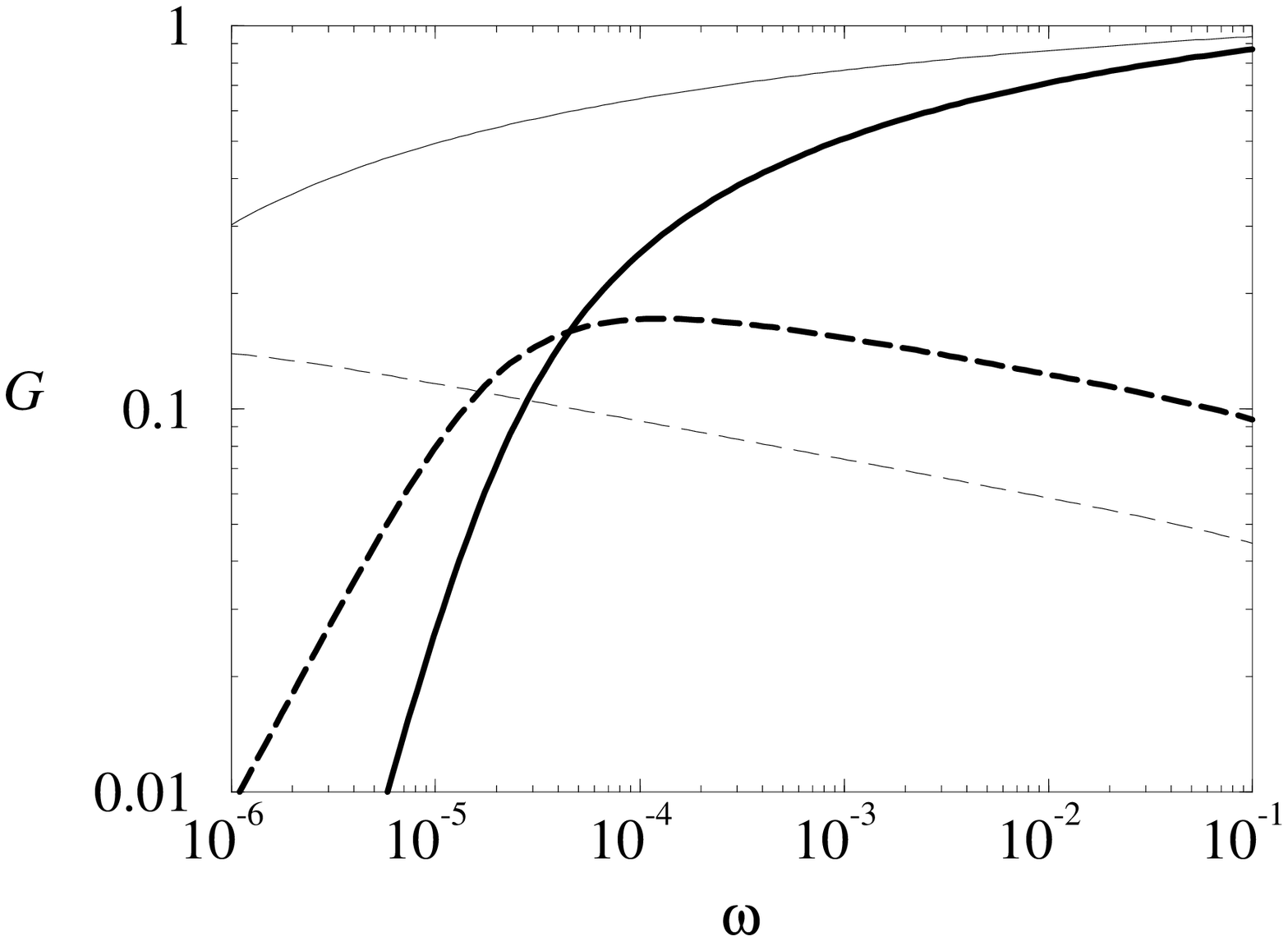}
\efig{moduli_with_cutoff}{Linear moduli $G'$ (solid line) and $G''$
(dashed) vs frequency $\om$ at $x=0.9$ with energy cutoff $\Emax=10$
(thick lines) and $\Emax=15$ (thin lines). The loss modulus increases as
$G''\sim \omega^{x-1}$ as the frequency decreases; at very low
frequencies, there is a cross-over to Maxwellian behavior.}
This applies as long as $\omega$ is still large compared to the cutoff
frequency, $\ommin=\exp(-\Emax/x)$. In this frequency regime, $G''$
therefore increases as $\omega$ decreases, again in qualitative
agreement with some recent experimental
observations~\cite{MasBibWei95,PanRouVuiLuCat96,HofRau93,MasWei95}.
An example is shown in Fig.~\ref{fig:moduli_with_cutoff}.

The above results relate to the ``equilibrium'' (pseudo) glass phase.
The time to reach this equilibrium state is expected to be of the
order of the inverse of the smallest relaxation rate,
$\ommin^{-1}=\exp(\Emax/x)$. For large $\Emax$, this may be much
larger than experimental time scales, and the non-equilibrium
behavior will then become relevant instead. We give only a brief discussion
here and refer to a future publication~\cite{aging_El} for more
details.  From the CE~(\ref{ce},\ref{ce_Gam}), it
can be deduced quite generally that the stress response to a small
oscillatory strain $\gam(t)=\gam\Re\exp(i\om t)$ switched on at $t=0$
is
\[
\sig(t)=\gam\Re\left[\Gcomp(\om, t) e^{i\om t}\right]
\]
with a time-dependent dynamic modulus
\be
\Gcomp(\om, t) = 1 - \int_0^t dt'\, e^{-i\om(t-t')}\,\G(t')\,\Grho(t-t')
\label{G_time_dep}
\ee
\bfig{9cm}{0cm}{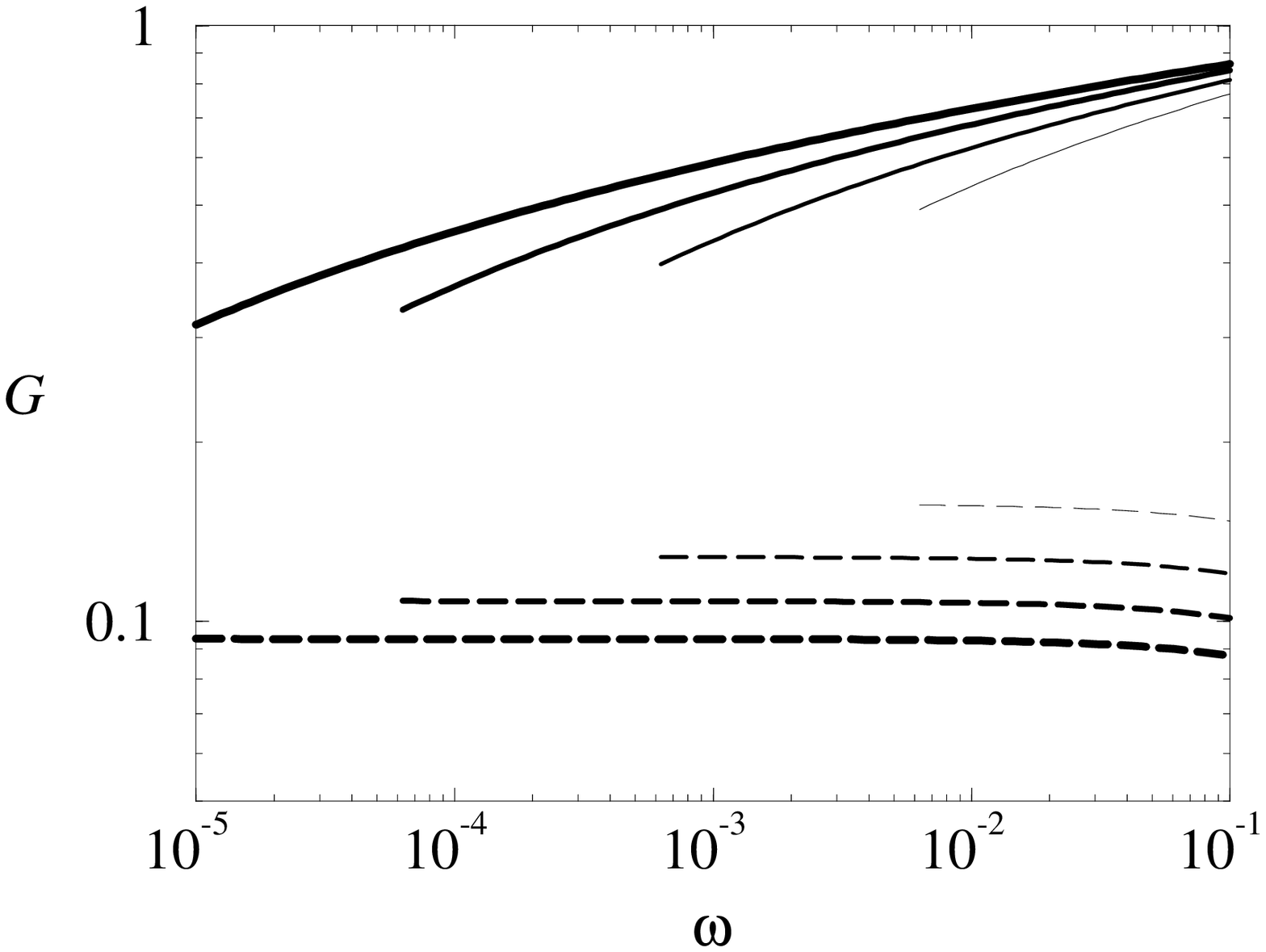}
\efig{G_time_dep}{Age-dependence of the dynamic moduli.
Shown are $G'$ (solid line) and $G''$
(dashed) vs frequency $\om$ at $x=1$; lines of increasing thickness
correspond to increasing age of the system: $t=10^4$, $10^5$, $10^6$,
$10^7$. Frequencies are restricted to the range $\om t \geq 2\pi\cdot
10$, corresponding to a measurement of $\Gcomp(\om, t)$ over at least
ten oscillation periods. Note the difference in horizontal and
vertical scales; both $G'$ and $G''$ have a very ``flat''
$\om$-dependence.}
This modulus is physically measurable only for $\om t$ significantly
greater than unity, of course, corresponding to a measurement over at
least a few periods.  Here we consider the case of an initial
distribution of yield energies $P_0(E)=\rho(E)$ (hence
$\Gzero\equiv\Grho$), corresponding to a ``quench'' at $t=0$ from
$x\to\infty$ to a finite value of $x$. We solve eq.~\(ce_Gam) for the
yielding rate $\G(t)$ numerically and then evaluate $\Gcomp(\om, t)$
using~\(G_time_dep).  Fig~\ref{fig:G_time_dep} shows the results for a
quench to the glass transition ($x=1$). Not unexpectedly, the
frequency dependence of the moduli follows the same power laws as in
the ``equilibrium'' glass discussed above; the amplitude of these,
however, depends on the ``age'' $t$ of the system. For $x<1$, one
finds $1-\Gcomp(\om, t)\sim (\om t)^{x-1}$~\cite{aging_El}; this
time dependence is the same as for the yielding rate
$\G(t)$~\cite{MonBou96}, and is closely related to the aging of the
susceptibility in Bouchaud's glass model~\cite{Bouchaud92}.  The
behavior of the loss modulus at the glass transition is particularly
noteworthy: Whereas $G''(\om,t)$ does tend to zero for $t\to\infty$,
it does so extremely slowly (as $1/\ln t$), while at the same time
exhibiting an almost perfectly ``flat'' ($G''\sim\omega^0$ for small
$\om$) frequency dependence. Where such an $\om$-dependence is
observed experimentally it may well, therefore, correspond to a
rheological measurement in an out-of-equilibrium, aging regime.  In
order to test this scenario directly, experiments designed to measure
a possible age dependence of the linear moduli would be extremely
interesting. Such experiments would obviously have to be performed on
systems where other sources of aging (such as coalescence in emulsions
and foams, evaporation of solvent etc) can be excluded; suspensions of
microgel beads, hard sphere colloids or colloid-polymer mixtures might
therefore be good candidates.

\subsection{Frustration}
\label{subsec:linear_frustration}

As pointed out in Sec.~\ref{subsec:model}, the SGR model in its basic
form~\(basic) assumes that after yielding, each element of a SGM relaxes
to a completely unstrained state, corresponding to a
local strain of $l=0$. This is almost certainly an oversimplification:
Frustration arising from interaction of an element with its neighbors
will in general prevent it from relaxing completely to its new
equilibrium state. This leads to a nonzero local strain $l$ directly
after yielding. This effect can be built into the model by replacing
the factor $\delta(l)$ in~\(basic) by a probability distribution
$q(l;E)$ of the local strain $l$ after yielding; this distribution
will in general also depend on the new yield energy $E$ of the element. We
consider here the case of ``uniform frustration'', where the strain $l$
after yielding has equal probability of taking on any value between
$-\lc$ and $\lc$, with $\lc=(2\E)^{1/2}$ being the typical yield
strain associated with the new yield energy. Because values of $l$ outside
this interval would not make much sense (the element would yield again
almost immediately), this scenario can be regarded as maximally frustrated.

\bfig{8cm}{0mm}{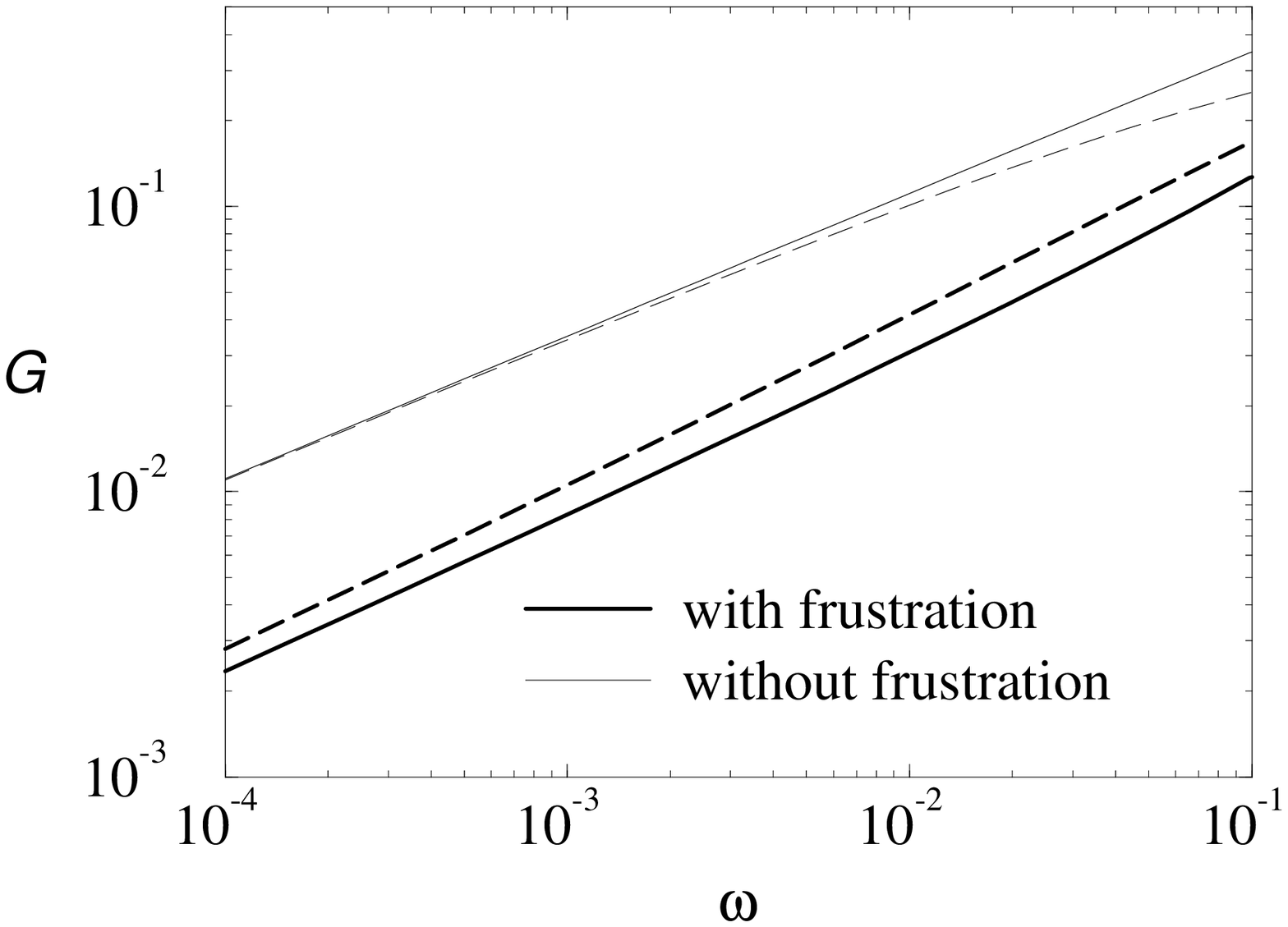}
\efig{moduli_with_frustration}{Effect of frustration. 
Shown are $G'$ (solid line) and $G''$
(dashed) vs frequency $\om$ at $x=1.5$; results for uniform
frustration (in bold) are compared with the unfrustrated case (thin
lines).}
An exact CE for such a frustrated scenario can
still be derived, but it is rather more cumbersome
than~(\ref{ce},\ref{ce_Gam}) due to extra integrations over the strain
variable $l$. The dynamic moduli, however, can still be worked out
fairly easily by considering a small perturbation around the steady
state of~\(basic) [with $\delta(l)$ replaced by $q(l;E)$]. One finds
\[
\Gcomp(\om) = \lav \frac{i\om\tau}{i\om\tau+1} + \frac{l^2}{x} 
\frac{i\om\tau}{(i\om\tau+1)^2} \rav\eq
\]
where the relaxation times $\tau=\exp[(E-\half l^2)/x]$ are now
dependent on both $E$ and $l$, and the equilibrium distribution over
which the average is taken is $P\eq(E,l)\propto\exp[(E-\half
l^2)/x]\rho(E)q(l;E)$. For the uniform frustration case, where $q(l;E) =
\Theta(E-\half l^2)/(8E)^{1/2}$, the dynamic moduli are compared with
the unfrustrated case in Fig.~\ref{fig:moduli_with_frustration}.
The main effect of frustration is to add a contribution to the
relaxation time spectrum near $\tau\approx 1$; this arises from
elements which have a strain $l\approx\pm\lc$ after yielding and
therefore yield again with a relaxation rate of order
unity. Otherwise, however, the main qualitative features of the
unfrustrated model are preserved; in particular, it can be shown that
the low frequency power law behavior~\(G_low_freq) remains unchanged.
We expect that the same will be true for other rheological properties
and therefore neglect frustration effects in the following.

\section{Nonlinear rheology}
\label{sec:nonlinear}

Arguably, the {\em linear} rheological behavior described in the
previous section follows inevitably from the existence of a power law
distribution of relaxation times. If we were only interested in the
linear regime, it would be simpler just to postulate such a power
law. The main attraction of the SGR model is, however, that it also
allows nonlinear rheological effects to be studied in detail. It is to
these that we now turn.

\subsection{Steady shear flow}
\label{subsec:steady_shear}

\subsubsection{Flow curves}

Steady shear flow ($\gamdot=$ const.) is one of the simplest probes of
nonlinear rheological effects. For the SGR model, the flow curve
(shear stress as a function of shear rate) can be calculated either
from the long-time limit of the CE~(\ref{ce},\ref{ce_Gam}), or
directly from the steady state solution of the equation of
motion~\(basic).  Either way, one obtains for the shear stress
\be
\sigma(\gamdot) = \frac{\int_0^\infty dl\ l\, \Grho(Z(l))}
{\int_0^\infty dl\ \Grho(Z(l))}
\label{flow_curve}
\ee
where
\be
Z(l)= \frac{1}{\gamdot} \int_{0}^l dl' \, e^{l'^2/2x}
\label{Zl_def}
\ee
\bfig{9cm}{0mm}{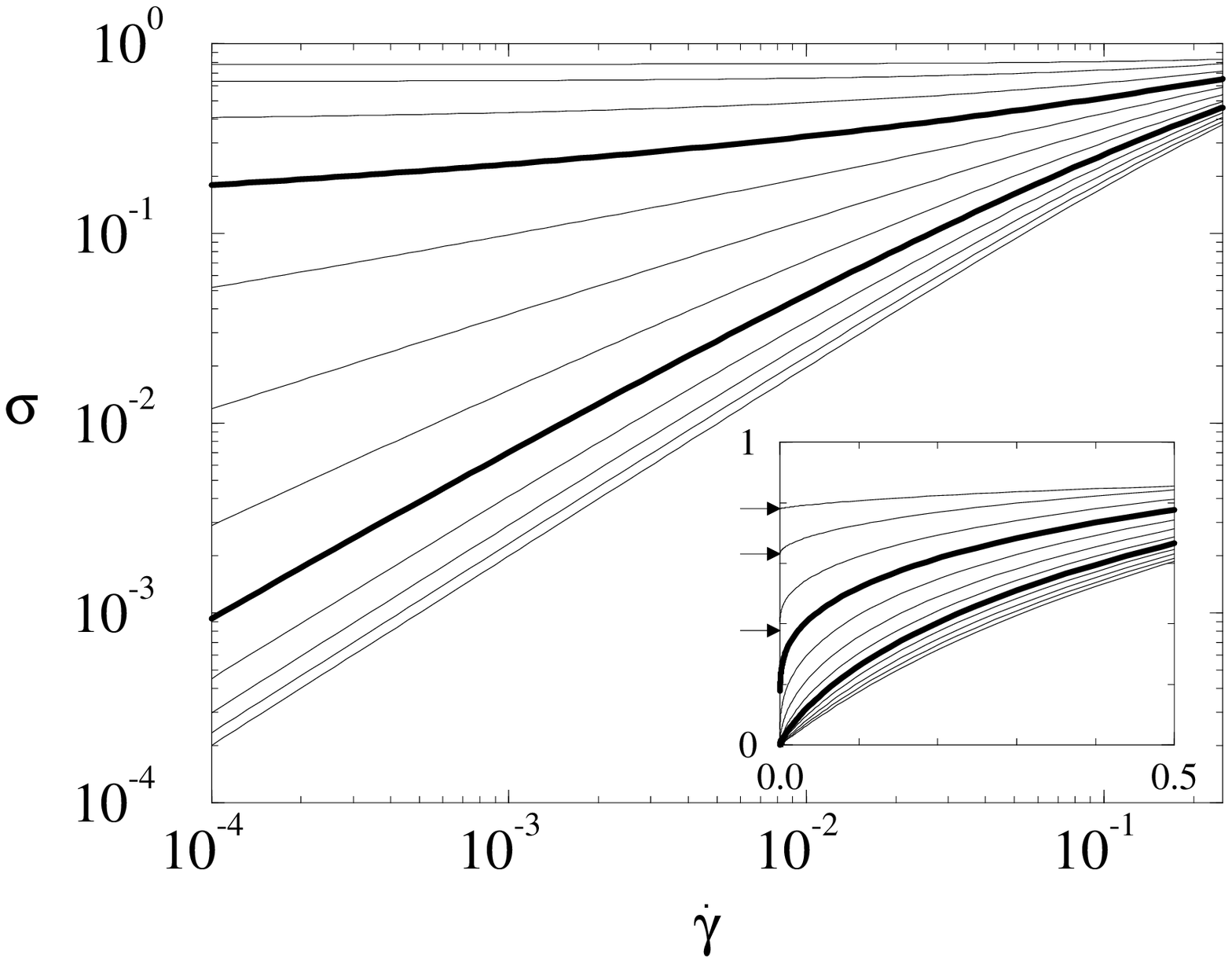}
\efig{flow_curves_combined}{Shear stress $\sig$ vs shear rate
$\gamdot$, for $x=0.25$, 0.5, $\ldots$, 2.5 (top to bottom on left);
$x=1$ and 2 are shown in bold~\protect\cite{num_error_footnote}. The inset
shows the behavior on a linear scale, with yield stresses for $x<1$
indicated by arrows.}
Eq.~\(flow_curve) is just the local strain averaged over its steady
state distribution, which is proportional to $\Grho(Z(l))$ (for
$l>0$).  The resulting stress can easily be evaluated numerically to
give the results in Fig.~\ref{fig:flow_curves_combined}.
For large shear rates $\gamdot\gtappr 1$, the
shear stress $\sig$ increases very slowly for all $x$
($\sig\sim(x\ln\gamdot)^{1/2}$), corresponding to strong shear
thinning. More interesting (and more physically
relevant~\cite{large_strain_rate_note}) is the small $\gamdot$
behavior, where we find three regimes:

(i) For $x>2$, the system is Newtonian, $\sig=\visc\gamdot$, for
$\gamdot\to 0$. The viscosity can be derived by noting that in this
regime, the size of the local strains $l$ that contribute
significantly to $\sig$ is proportional to $\gamdot$. For $\gamdot\to
0$, it decreases to zero, and we can approximate $Z(l)=l/\gamdot$,
giving
\bea
\eta &=& \frac{\sig}{\gamdot}=
\frac{\int_0^\infty dt\ t\, \Grho(t)} {\int_0^\infty dt\ \Grho(t)}
\nonumber\\
&=& \G\eq\int dE\ \rho(E)\; e^{2E/x} = \lav e^{E/x} \rav\eq 
= \lav \tau \rav\eq
\nonumber
\eea
The viscosity is therefore simply the average of the relaxation time
$\tau=\exp(E/x)$ over the equilibrium distribution of energies,
$P\eq(\E)=\G\eq\exp(\E/x)\rho(\E)$. From the form
$\visc\propto\lav\exp(2\E/x)\rav_\rho$ one sees that it diverges at
$x=2$, \ie, at {\em twice} the glass transition temperature. The
existence of several characteristic temperatures in the SGR model is
not surprising; in fact, Bouchaud's original glass model already has
this property~\cite{MonBou96} (which has also been discussed in more
general contexts, see \eg~\cite{Odagaki95}).

(ii) The divergence of the viscosity for $x\to 2$ signals the onset of
a new flow regime: for $1<x<2$ one finds power law fluid rather than
Newtonian behavior. The power law exponent can be derived as follows:
The steady shear stress~\(flow_curve) is the ratio of the integrals
\[
I_n(\gamdot)=\int_0^\infty dl\ l^n\,\Grho(Z(l))
\]
for $n=1$ and $n=0$. By techniques very similar to those used in
App.~\ref{app:Grho_asympt}, one derives that in the small $\gamdot$
limit, $I_n$ scales as $\gamdot^{n+1}$ for $x>n+1$; for lower $x$,
there is an additional contribution scaling as $\gamdot^x$ up to sub-power
law factors (see App.~\ref{app:yield_stress}). The dominant
contribution to $\sig$ for small $\gamdot$ in the regime $1<x<2$
therefore scales as $\sig\sim\gamdot^{x-1}$, again up to sub-power law
factors. The power law fluid exponent therefore decreases linearly,
from a value of one for $x=2$ to zero at the glass transition $x=1$.

(iii) For $x<1$, the system shows a yield stress: $\sig(\gamdot\to
0)=\sigy>0$. This can again be understood from the scaling of $I_1$
and $I_0$: the dominant small $\gamdot$ contributions to both scale as
$\gamdot^x$ for $x<1$, giving a finite ratio $\sigy=I_1/I_0$ in the
limit $\gamdot\to 0$. For general $\rho(E)$ there are
subtleties due to sub-power law corrections here, which are discussed
in App.~\ref{app:yield_stress}. Here we focus on the simplest
case~\(exp_rho) of exponential $\rho(E)$, where such corrections are
absent. Using the scaling of $I_1$ and $I_0$, we can then write the
shear stress for small $\gamdot$ as
\be
\sig=
\frac{O(\gamdot^x)+O(\gamdot^2)}{O(\gamdot^x)+O(\gamdot^1)}=
\sigy+O(\gamdot^{1-x})
\label{HB_exp_rho}
\ee
Beyond yield, the stress therefore again increases as a power law of
the shear rate, $\sig-\sigy\propto \gamdot^{1-x}$. For exponential
$\rho(E)$, the yield stress itself can be calculated explicitly: In
order to have $\sigy>0$, the values of $l$ that contribute to the
shear stress~\(flow_curve) must remain finite for $\gamdot\to 0$. But
then for any fixed $l$, $Z(l)\to\infty$. We can therefore use the
asymptotic form $\Grho(z)=x! z^{-x}$ in~\(flow_curve), giving
\be
\sigy= \frac{\int_0^\infty dl\ l\, [Z(l)]^{-x}}
{\int_0^\infty dl\ [Z(l)]^{-x}}
\label{sigy_exp_rho}
\ee
\bfig{8cm}{0mm}{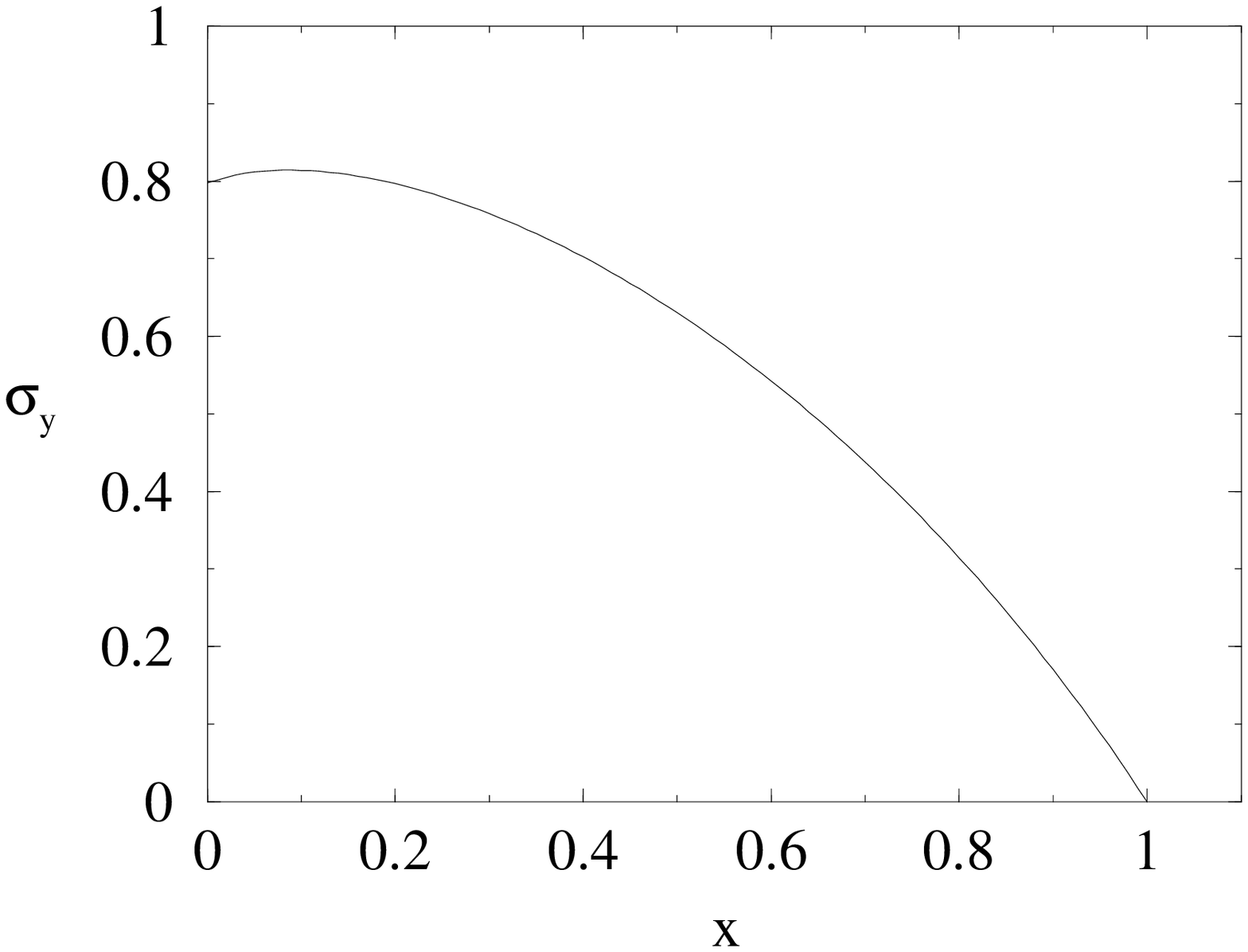}
\efig{yield_stress}{Yield stress $\sigy$ as a function of $x$.}
The factors $\gamdot^x$ (from the definition~\(Zl_def) of $Z(l)$) in
numerator and denominator have canceled, making the result independent of
$\gamdot$ as required.  Fig.~\ref{fig:yield_stress} shows the
resulting yield stress as a function of $x$; it has a linear onset
near the glass transition, $\sigy\sim 1-x$.

To summarize, the behavior of the SGR model in regimes (ii) and (iii)
matches respectively the power-law fluid
~\cite{Holdsworth93,Dickinson92,BarHutWal89} and
Herschel-Bulkeley~\cite{Holdsworth93,Dickinson92} scenarios as used to
fit the nonlinear rheology of pastes, emulsions, slurries, etc. In
regime (ii), the power law exponent is simply $x-1$, $x$ being the
effective (noise) temperature; in regime (iii) and for exponential
$\rho(E)$, it is $1-x$ (see App.~\ref{app:yield_stress} for a
discussion of the general case).  Numerical data for the effective
exponent $d\ln(\sig-\sigy)/d\ln\gamdot$ in
Fig.~\ref{fig:flow_curves_eff_exponents_combined} are compatible with
this, although the exponent only approaches its limiting value very
slowly as $\gamdot\to 0$ for $x$ near the boundaries of the power law
regime, $x=1$ and 2.
\bfig{8cm}{0mm}{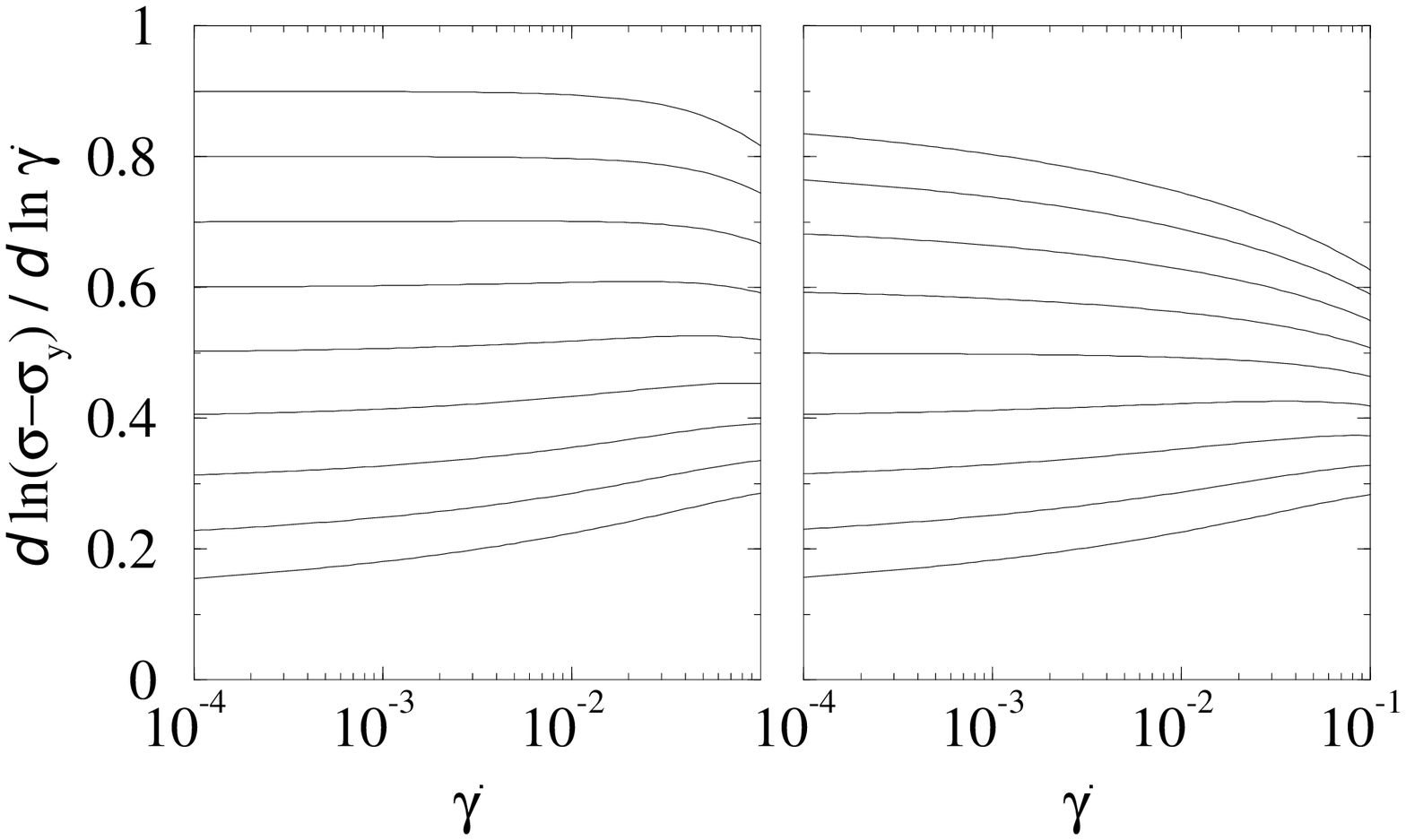}
\efig{flow_curves_eff_exponents_combined}{Effective power law exponent
$d\ln(\sig-\sigy)/d\ln\gamdot$ vs $\gamdot$ in the glass phase (left,
yield stress $\sigy>0$, $x=0.1$, 0.2 \ldots 0.9 from top to bottom)
and in the power law fluid regime (right, $\sigy=0$, $x=1.1$, 1.2
\ldots 1.9 from bottom to top).}

A natural question to ask is of course how the existence of a yield
stress in the glass phase affects the linear moduli, \ie, the response
to small strains. This is a highly nontrivial issue due to the
non-ergodicity of the glass phase and the corresponding aging
behavior. In particular, the answer will depend to a significant
degree on the strain history of the material. We therefore leave this
point for future, more detailed study~\cite{aging_El}.

\subsubsection{Flow interrupts aging}

We saw above that there is a steady state regime for {\em any value
of} $x$ in the presence of steady shear flow.  On the other hand, the
discussion in Secs.~\ref{sec:ce} and~\ref{subsec:linear_glass} showed
that in the absence of flow, the system has no steady state in the
glass phase ($x<1$) and instead exhibits aging behavior.  The
difference between the two cases can be seen more clearly by
considering the distribution of yield energies, $P(E)$.  Without flow,
one obtains a Boltzmann distribution $P(\E)\propto\rh(\E)\exp(\E/x)$
up to (for $x<1$) a ``soft'' 
cutoff which shifts to higher and higher energies
as the system ages~\cite{MonBou96}. This cutoff, and hence the most
long-lived traps visited (which have a lifetime comparable to the age
of the system), dominate the aging behavior~\cite{Bouchaud92}. In the
presence of flow, on the other hand, there is a finite steady state
value for this cutoff; one finds
\bea
P(\E) & \propto & \rh(\E)\;e^{\E/x} \; \mbox{\ \ for\ \,}
\E \ll x\ln(\gamdot^{-1}x^{1/2}) \nonumber\\
P(\E) & \propto & \rh(\E)\;\E^{1/2} \: \mbox{\ \ for\ \,}
\E \gg x\ln(\gamdot^{-1}x^{1/2})
\label{PE_steady_shear}
\eea
(only the second regime exists for $\gamdot\gtappr x^{1/2}$).  The
existence of these two regimes can be explained as follows: Assume the
yielding of an element is noise-induced. Its typical lifetime is then
$\exp(E/x)$, during which it is strained by $\gamdot\exp(E/x)$. The
assumption of noise-induced yielding is self-consistent if this amount
of strain does not significantly enhance the probability of yielding,
\ie, if $[\gamdot\exp(E/x)]^2/x \ll 1$. This is the low $E$ regime
in~\(PE_steady_shear), which gives a Boltzmann form for the yield
energy distribution as expected for noise-induced yielding. In the
opposite regime, yielding is primarily strain induced, and the time
for an element to yield is of the order of
$\lc/\gamdot=(2E)^{1/2}/\gamdot$ (rather than $\exp(E/x)$).
Intuitively, we see that flow prevents elements from getting stuck in
progressively deeper traps and so truncates the aging process after a
finite time. We can therefore say that ``flow interrupts
aging''~\cite{BouDea95}.

\subsubsection{Cox-Merz rule}

A popular way of rationalizing flow curves is by relating them to the
linear rheology via the heuristic Cox-Merz rule~\cite{CoxMer58}. This
rule equates the ``dynamic viscosity''
$\dynvisc(\om)=|\Gcomp(\om)|/\om$ with the steady shear viscosity
$\visc(\gamdot)=\sig(\gamdot)/\gamdot$ when evaluated at
$\gamdot=\om$. The ratio $\om\visc(\gamdot=\om)/|\Gcomp(\om)|$ is
therefore equal to unity if the Cox-Merz rule is obeyed
perfectly. Using our previous results, we can easily verify whether
this is the case in the SGR model.  From Fig.~\ref{fig:cox_merz}, we
see that in the Newtonian regime $x>2$, the Cox-Merz rule is obeyed
reasonably well for frequencies $\om\ltappr 1$; for $\om\to 0$, it
holds exactly as expected (recall that $\eta(\gamdot)=\lav\tau\rav$,
while from~\(G_low_freq), $\Gcomp(\om\to 0)=i\om\lav\tau\rav$). In the
power-law fluid regime $1<x<2$, on the other hand, the Cox-Merz rule
is seen to be less reliable and is not obeyed exactly even in the zero
frequency limit. At the glass transition ($x\to 1$), it fails rather
dramatically: In this limit, $|\Gcomp(\om)|=1$ and so the Cox-Merz
rule predicts a shear rate independent shear stress
$\sig(\gamdot)=\gamdot\visc(\gamdot)=1$, whereas in fact
$\sig(\gamdot)$ decreases to zero for $\gamdot\to 0$.
\bfig{9cm}{0mm}{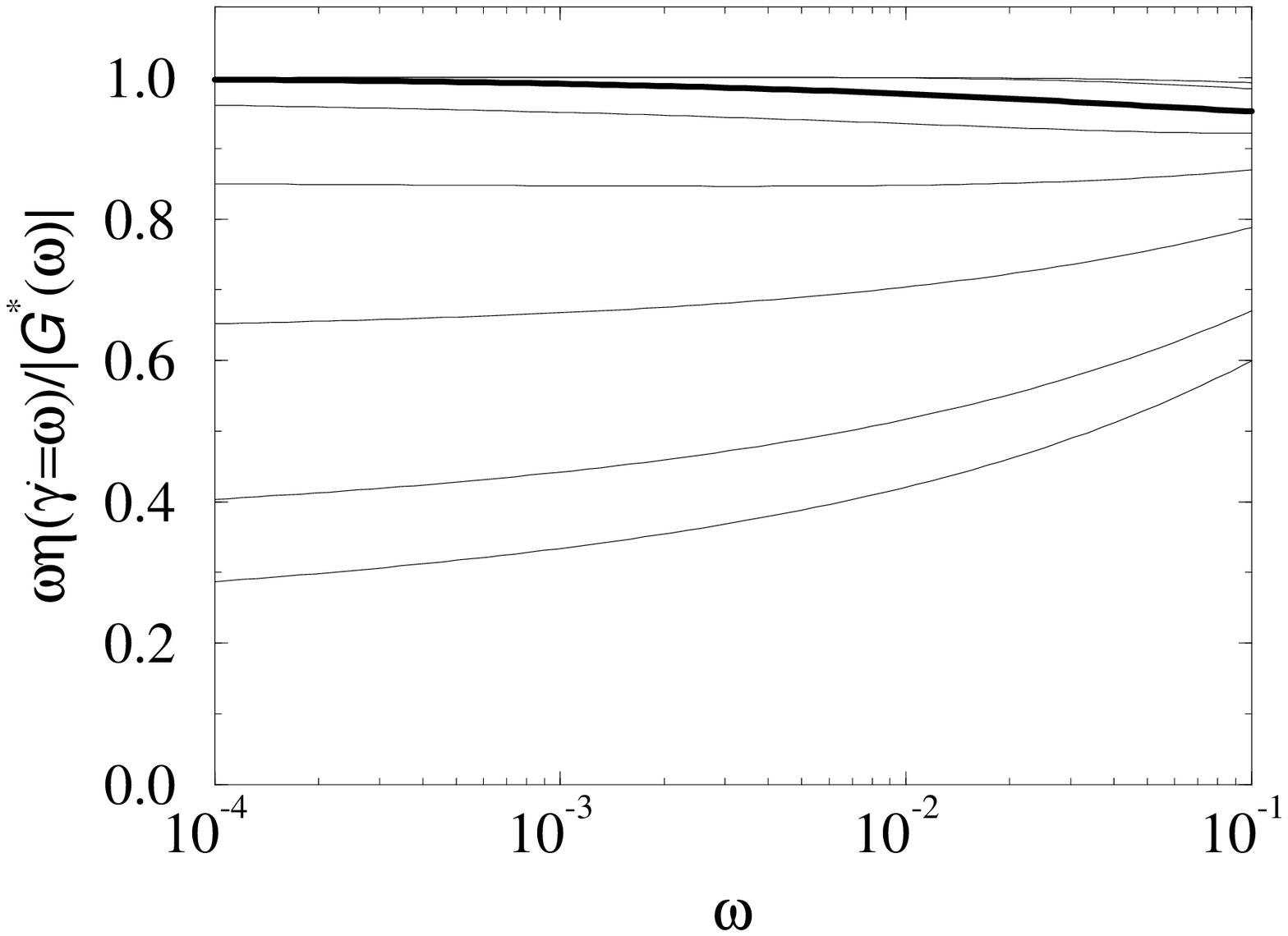}
\efig{cox_merz}{Cox-Merz ratio $\om\visc(\gamdot=\om)/|\Gcomp(\om)|$
as a function of $\om$ for noise temperatures $x=1,$ 1.2, 
\ldots, 1.8, 2 (bold), 2.5, 3 (bottom to top).}

\subsubsection{Dissipation under steady shear}

Finally, in conclusion of this section on steady shear flow, we
calculate the distribution of energies dissipated in yield events.
This distribution may provide a useful link to computer simulations of
steady shear flow of foams, for example, where it is often easy to
monitor discontinuous drops in the total energy of the system and
determine their distribution~\cite{Durian97}. The correspondence is,
however, not exact. Our mean-field model treats all yield events as
uncorrelated with each other, both in time and space. In reality, such
correlations will of course exist. In fact, several events may occur
simultaneously, at least within the time resolution of a simulation or
experiment. The observed drop in total energy would then have to be
decomposed into the contributions from the individual events to allow
a direct comparison with our model. This is only possible if the
events are sufficiently localized (spatially) to make such a
decomposition meaningful. In foams and emulsions, there is evidence
that this may indeed be the
case~\cite{LacGreLevMasWei96,Durian97,HebLeqMunPin97,%
LiuRamMasGanWei96,HutWeaBol95,GopDur95,DurWeiPin91,EarWil96}.

We earlier derived the energy balance equation~\(energy_balance) and
deduced from it that, within the model, each yield event dissipates
the elastic energy $\de=\half l^2$ stored in the element just prior to
yielding. The probability of observing a yield event with energy
dissipation $\de$ is therefore given by
\[
P(\de)=\frac{1}{\G}
\int \! dE\, dl \ P(E, l)\, e^{-(E-\half l^2)/x}\, \delta\!
\left(\de-\half l^2\right)
\]
The steady state distribution $P(E,l)$ of yield energies and
local strains for a given shear rate $\gamdot$ and noise temperature
$x$ can easily be deduced from~\(basic). After some algebra, the
result can be put into the simple form
\[
P(\de)\, d\de = - \deriv{l}\; \Grho(Z(l))\, dl
\]
Fig.~\ref{fig:diss_distr} shows the resulting $P(\de)$ for exponential
$\rho(E)$. Larger shear rates $\gamdot$ are seen to lead to an
increasing dominance of ``large'' yield events (which dissipate a lot
of energy). This is intuitively reasonable: the larger $\gamdot$, the
larger the typical strains of elements when they yield. The functional
dependence of $P(\de)$ on $\de$ is surprisingly simple. An initial
power law decay $P(\de)\sim\de^{-1/2}$ crosses over for
$\de\approx\gamdot^2$ into a second power law regime
$P(\de)\sim\de^{-1-x/2}$. This is cut off exponentially for values of
$\de$ around unity~\cite{P_de_bump_note}. The exponential tail for
very large dissipated energies is $P(\de)\sim\exp(-\de)$ independently
of $x$. This asymptotic behavior is the same as for the prior density
of yield energies, $\rho(E)\sim\exp(-E)$; measurements of $P(\de)$ for
large $\de$ could therefore yield valuable information on $\rho(E)$.
\bfig{8cm}{0cm}{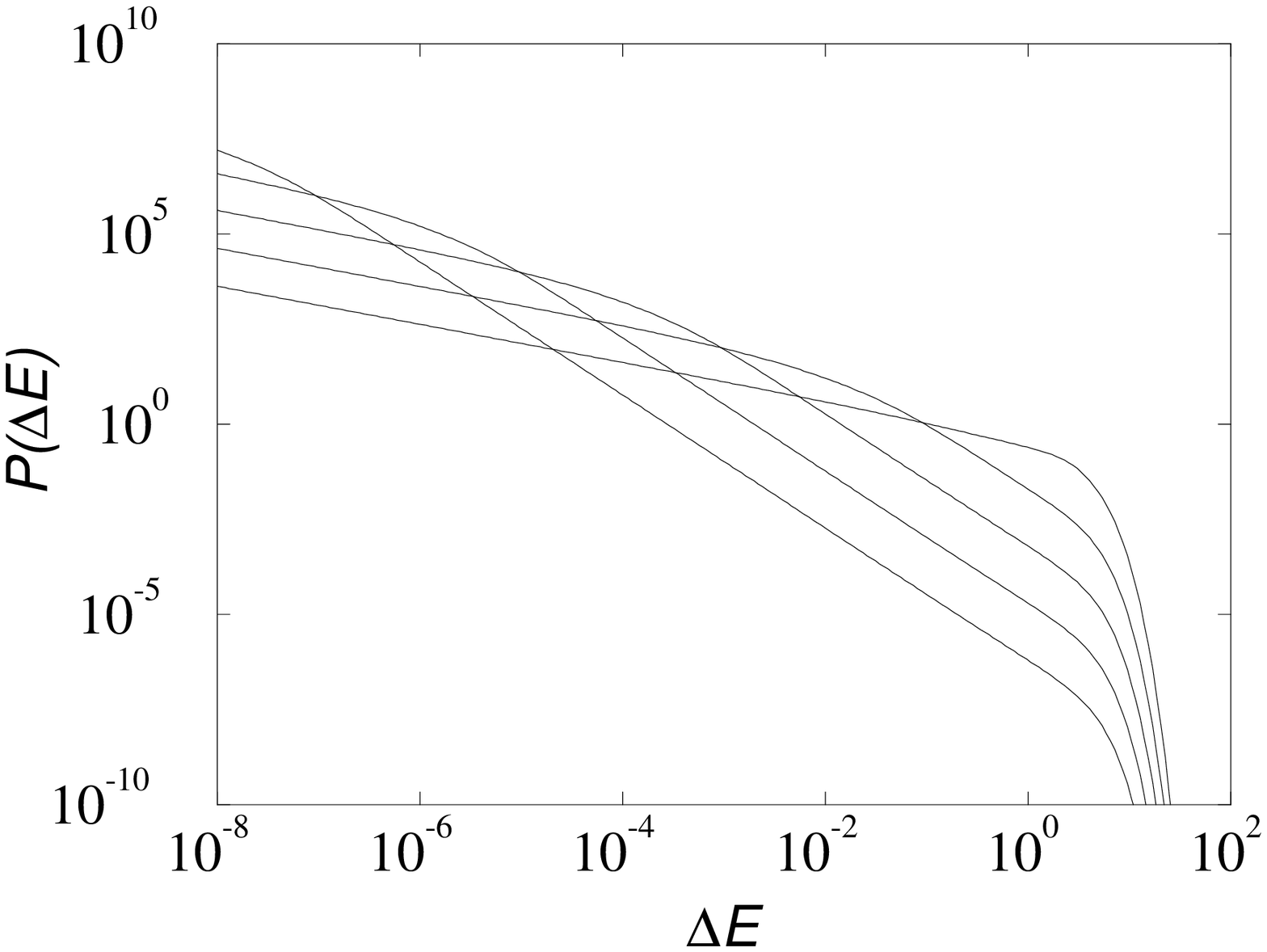}
\efig{diss_distr}{Distribution $P(\de)$ of energies $\de$ dissipated
in yield events under steady flow, for $x=1.5$ and $\gamdot=10^{-4}$,
$10^{-3}$, \ldots, 1 (bottom to top at $\de=1$)}

These results for $P(\de)$ also help to understand the small $\gamdot$
scaling of the energy dissipation rate
$\sig\gamdot=\G\lav\de\rav$. From the results of
Sec.~\ref{subsec:steady_shear}, we know that this is $\gamdot^2$ in
the Newtonian regime $x>2$, $\gamdot^x$ in the power law fluid range
$1<x<2$, and $\gamdot$ in the yield stress regime $x<1$.  (The limit
$\gamdot\to 0$ is always understood here and in the following.) The
form of $P(\de)$ suggests to decompose the dissipation into its
contributions from ``{\underline s}mall'' ($\de=O(\gamdot^2)$) and
``{\underline l}arge'' ($\de=O(1)$) dissipation events. Each of these
two classes makes a contribution to $\sig\gamdot$ which is the
fraction of elements in the class, times the average yielding rate in
the class, times the average energy dissipated. Hence, in obvious
notation,
\[
\sig\gamdot = P_s{\G}_s{\de}_s +
P_l{\G}_l{\de}_l
\]
One then easily confirms the following scalings.  The average {\em
dissipated energies} are obviously given by
${\de}_s=O(\gamdot^2)$ and ${\de}_l=O(1)$. The
average {\em yielding rate} for the small, noise induced events is
independent of shear rate, ${\G}_s=O(\gamdot^0)$; while for
the large, shear induced events it is
${\G}_l=O(\gamdot)$. Finally, for the {\em fractions} of
small and large elements, one finds that {\em above} the glass
transition, almost all elements have small strains $l=O(\gamdot)$,
corresponding to $\de=O(\gamdot^2)$; hence $P_s=O(1)$. Large strains,
on the other hand, occur with a probability $P_l=O(\gamdot^{x-1})$
which becomes vanishingly small for small shear rates. {\em Below} the
glass transition, the situation is reversed: $P_l=O(1)$, while
$P_s=O(\gamdot^{1-x})$. Putting everything together, one has:

(i) In the Newtonian regime ($x>2$), dissipation is dominated by
small, noise induced events, and is therefore of $O(\gamdot^2)$.

(ii) In the power law fluid range ($1<x<2$), a vanishingly small
number of elements has large strains, but these dominate the
dissipation $\sig\gamdot=P_l{\G}_l{\de}_l$ =
$O(\gamdot^{x-1})O(\gamdot)$ = $O(\gamdot^x)$. As the glass transition
is approached, the fraction of large elements and hence the
dissipation increases.

(iii) In the yield stress regime, most elements have large
strains, giving a dissipation rate $\sig\gamdot$ = $O(\gamdot)$ which
simply scales with the shear rate.
 
With the same approach, one can also analyse the total {\em yielding
rate} $\G=P_s{\G}_s+P_l{\G}_l$. Small events always dominate, and $\G$
therefore scales with $\gamdot$ in the same way as $P_s$. This is true
even in the non-Newtonian flow regimes ($x<2$), where the contribution
of these elements to the total {\em dissipation rate} is negligible.

The distribution of total energy drops $\Delta E_{\rm tot}$ due to
rearrangements has been monitored in recent simulations of steady
shear flow of two-dimensional foam, based on a ``soft-sphere
model''~\cite{Durian95,Durian97}. It was found to exhibit a power law
$P(\Delta E_{\rm tot}) \sim \Delta E_{\rm tot}^{-\nu}$ with an
exponent $\nu\approx 0.7$, with an exponential cutoff for large energy
drops. More recent simulations using the same model suggest that, when
$\Delta E_{\rm tot}$ is normalized by the average elastic energy per
foam bubble, the form of $P(\Delta E_{\rm tot})$ is largely
insensitive to variations in shear rate $\gamdot$. Decreasing the gas
volume fraction $\phi$, on the other hand, moves the (normalized)
cutoff to larger energies, suggesting a possible divergence near the
rigidity loss transition at $\phi\approx
0.64$~\cite{recent_soft_sphere_stuff}.
Simulations using a ``vertex
model'', on the other hand, gave $P(\de_{\rm tot})\sim \de_{\rm
tot}^{-3/2}$ with no system-size independent cutoff for large
$\de_{\rm tot}$~\cite{OkuKaw95}. It is unclear how these results can
be reconciled; neither, however, is fully compatible with the
predictions of the SGR model for $P(\de)$. At this point, we do not
know whether this disagreement is due to the difference between $\de$
(dissipation in a single yield event) and $\de_{\rm tot}$ (total
dissipation in a number of simultaneous yield events), or whether it
points to a more fundamental shortcoming of the SGR model such as
neglect of spatial or temporal correlations.

\subsection{Shear startup}

If a shear flow is started up at $t=0$, such that $\gam(t)=\gamdot t$
for $t\geq 0$, then $\sig(\gamdot)$ as given by the flow curve is the
asymptotic, steady state value of the stress for $t\to\infty$. We now
consider the transient behavior $\sig(t)$ for finite $t$. This depends
on the initial state of the system at $t=0$; here we consider only the
case where this initial state is the equilibrium state~\(eq_initial)
at the given value of $x$. This restricts our discussion to the regime
above the glass transition, $x>1$, where such an equilibrium state
exists~\cite{shear_startup_Emax_note}.  Solving the
CE~(\ref{ce},\ref{ce_Gam}) numerically, we can find the stress $\sig$
as a function of time $t$ or, alternatively, strain
$\gam$. Fig.~\ref{fig:shear_startup} shows exemplary results. The
initial behavior under shear startup is found to be elastic in all
cases, $\sig=\gam$. (This can in fact be deduced directly by
expanding~\(ce) to first order in $t$ and noting that
$\Gzero(\ZZ{t}{0})=1+O(t)$ while the contribution from the integral is
of $O(t^2)$.)  Asymptotically, on the other hand, the stress
approaches the steady-state (flow curve) value
$\sig(\gamdot)$. However, the model predicts that it does not
necessarily do so in a monotonic way. Instead, the stress can
``overshoot''; within the model, this effect is most pronounced near
the glass transition ($x\approx 1$). Such overshoot effects have been
observed experimentally in, for example, foam
flow~\cite{KhaSchneArm88}.  The tendency towards large overshoots for
$x\to 1$ agrees with our results for the linear moduli and flow
curves: As the glass transition is approached, the behavior of the
system becomes predominantly elastic; the stress can therefore
increase to larger values in shear startup before the material (as a
whole) yields and starts to flow.
\bfig{9cm}{0cm}{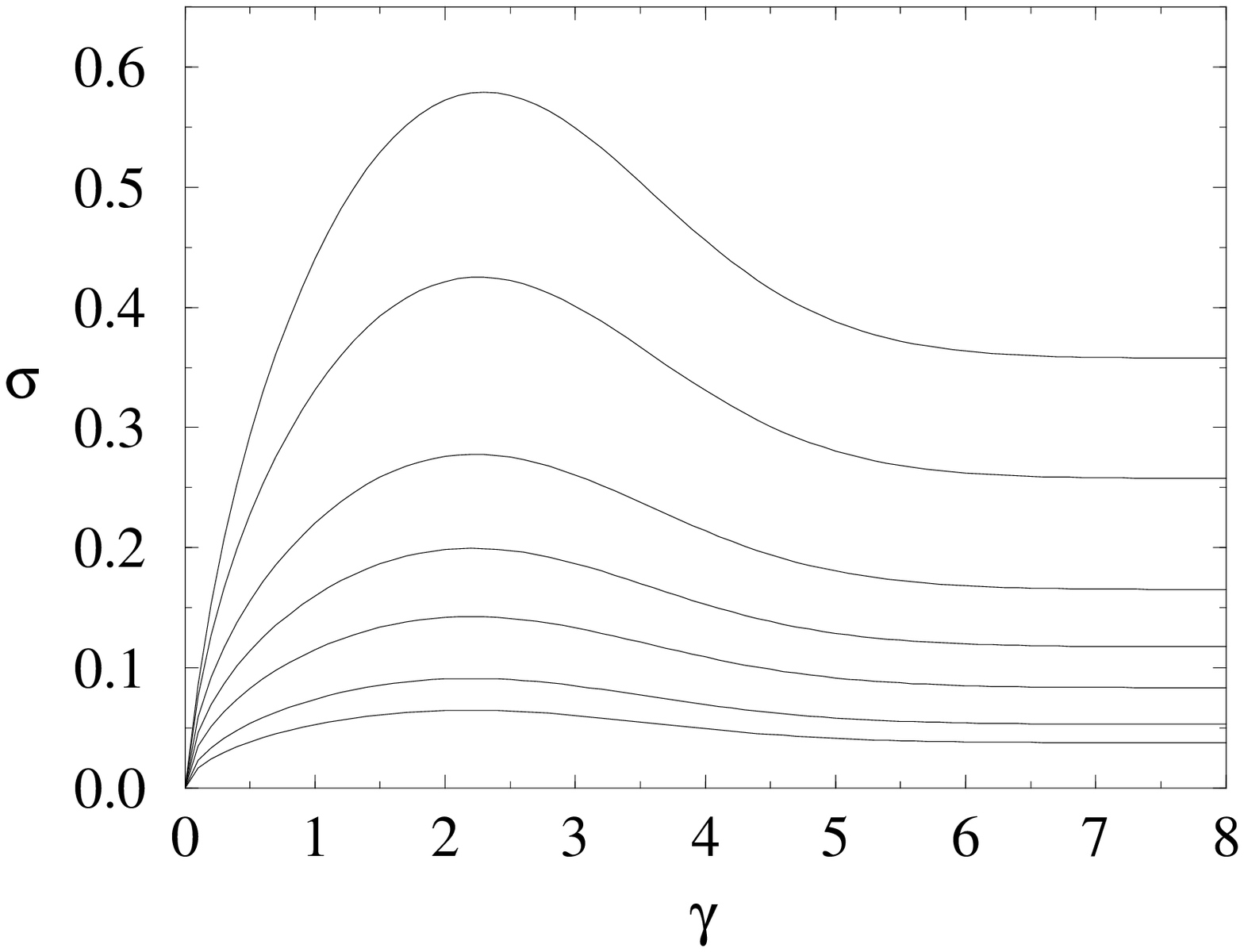}
\efig{shear_startup}{Stress $\sig$ vs strain $\gam$ for shear startup
at effective temperature $x=1.5$. The shear rate
$\gamdot=0.001$, 0.002, 0.005, 0.01, 0.02, 0.05, 0.1 increases from
bottom to top.}

\subsection{Large step strains}

As a further probe of the nonlinear rheological behavior predicted by
the SGR model, we now consider large (single and double) step
strains. Again, we do not discuss aging effects here and therefore
limit ourselves to the regime $x>1$ with the equilibrium initial
condition~\(eq_initial).

The case of a single step strain ($\gam(t)=\gam\Theta(t)$, with
$\Theta(t)=1$ for $t>0$ and zero otherwise) is particularly
simple. The integral over $t'$ in the CE~\(ce) is
then identically zero, giving a stress response of
\be
\sig(t)=\gam\Gzero(\ZZ{t}{0})=\gam G\eq\left(e^{\gam^2/2x} t\right)
\label{nonlinear_single_step}
\ee
Comparing with the response~\(lin_response) in the linear regime, the
effect of nonlinearity is to speed up all relaxation processes by a
factor $\exp(\gam^2/2x)$. It is easy to see why this is the
case. Because we are starting from an unstrained equilibrium
configuration, each element initially has $l=0$ and a yielding rate
$\exp(-E/x)$. Directly after the strain is applied, it therefore has
local strain $l=\gam$; this increases its relaxation rate to
$\exp[-(E-\half \gam^2)/x]$, \ie, by the same factor $\exp(\gam^2/2x)$
for all elements. Fig.~\ref{fig:single_steps} illustrates this effect
of strain nonlinearity; note that the stress for large step strains
can decay to small values faster than for small strains, due to the
strain-induced speed-up of all relaxation processes.
\bfig{9cm}{0cm}{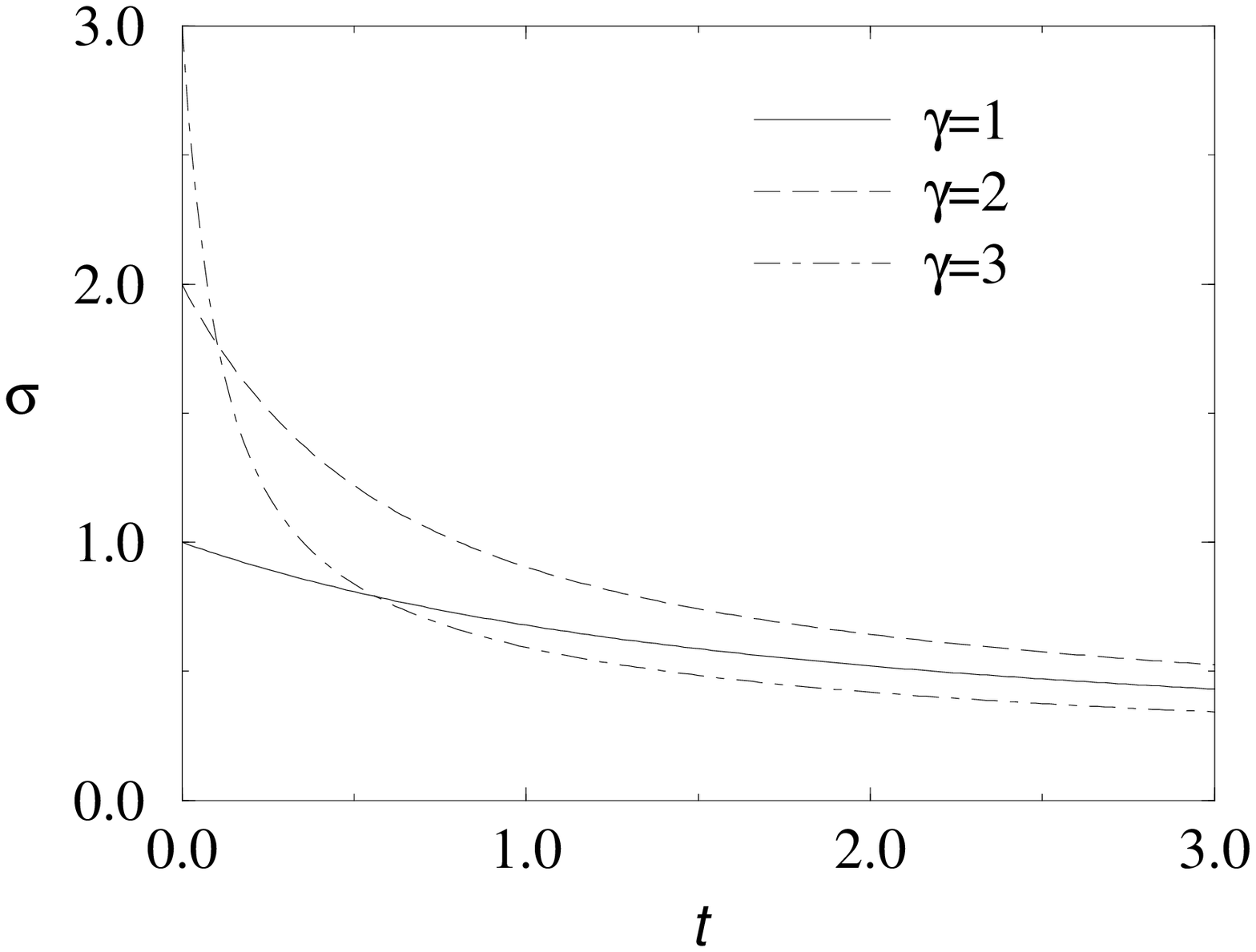}
\efig{single_steps}{Stress response to step strains of amplitude
$\gam=1$, 2, 3, at noise temperature $x=1.5$. 
}

Interestingly, the {\em instantaneous} response is always elastic and
not affected by nonlinear effects: $\sig(t=0^+)=\gam$ for all
$\gam$. It is easily shown from the CE~(\ref{ce},\ref{ce_Gam}) that
this is a general feature of the SGR model; whenever the macroscopic
strain $\gam(t)$ changes discontinuously by $\Delta\gam$, the stress
$\sig(t)$ changes by the same amount. We also note that the stress
response~\(nonlinear_single_step) cannot be factorized into time and
strain dependence. However, for the particular case of exponential
$\rh(\E)$ and long times $\exp(\gam^2/2x)t \gg 1$, such a
factorization does exist due to the asymptotic behavior of $G\eq$,
$G\eq(z)\sim z^{1-x}$. (This follows from $\Grho(z)\sim z^{-x}$
and~\(Grho_Geq).)  One then has
\[
\sig(t) \sim \gam h(\gam)G\eq(t) \qquad h(\gam)=\exp\left[
-\half(1-x^{-1})\gam^2\right]
\]
The product $\gam h(\gam)$ tends to zero as $\gam$ increases,
corresponding to a pronounced shear-thinning effect.

\bfig{9cm}{0cm}{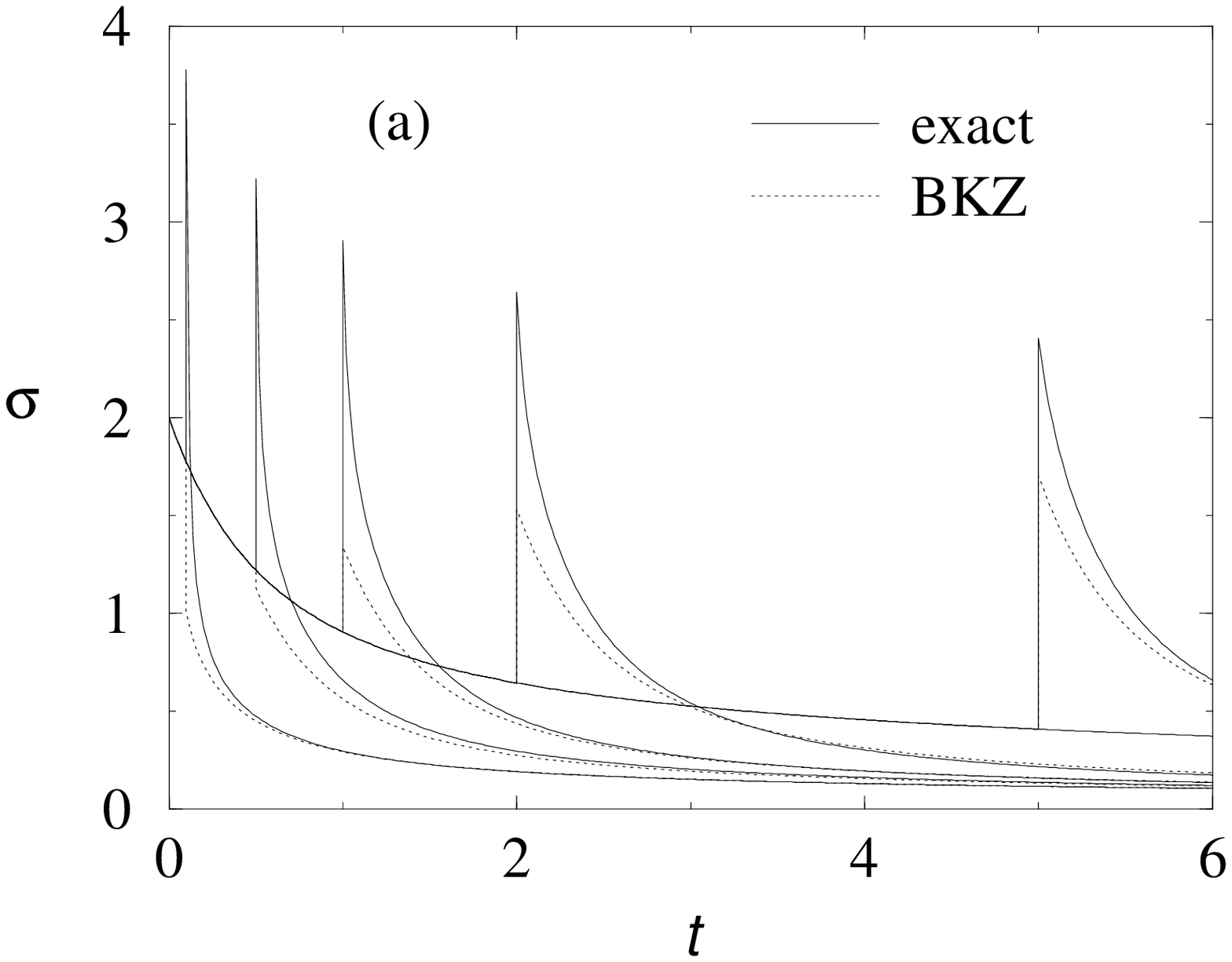}
\end{center}

\vspace*{-1cm}
\begin{center}
\epsfxsize 9cm
\epsfbox{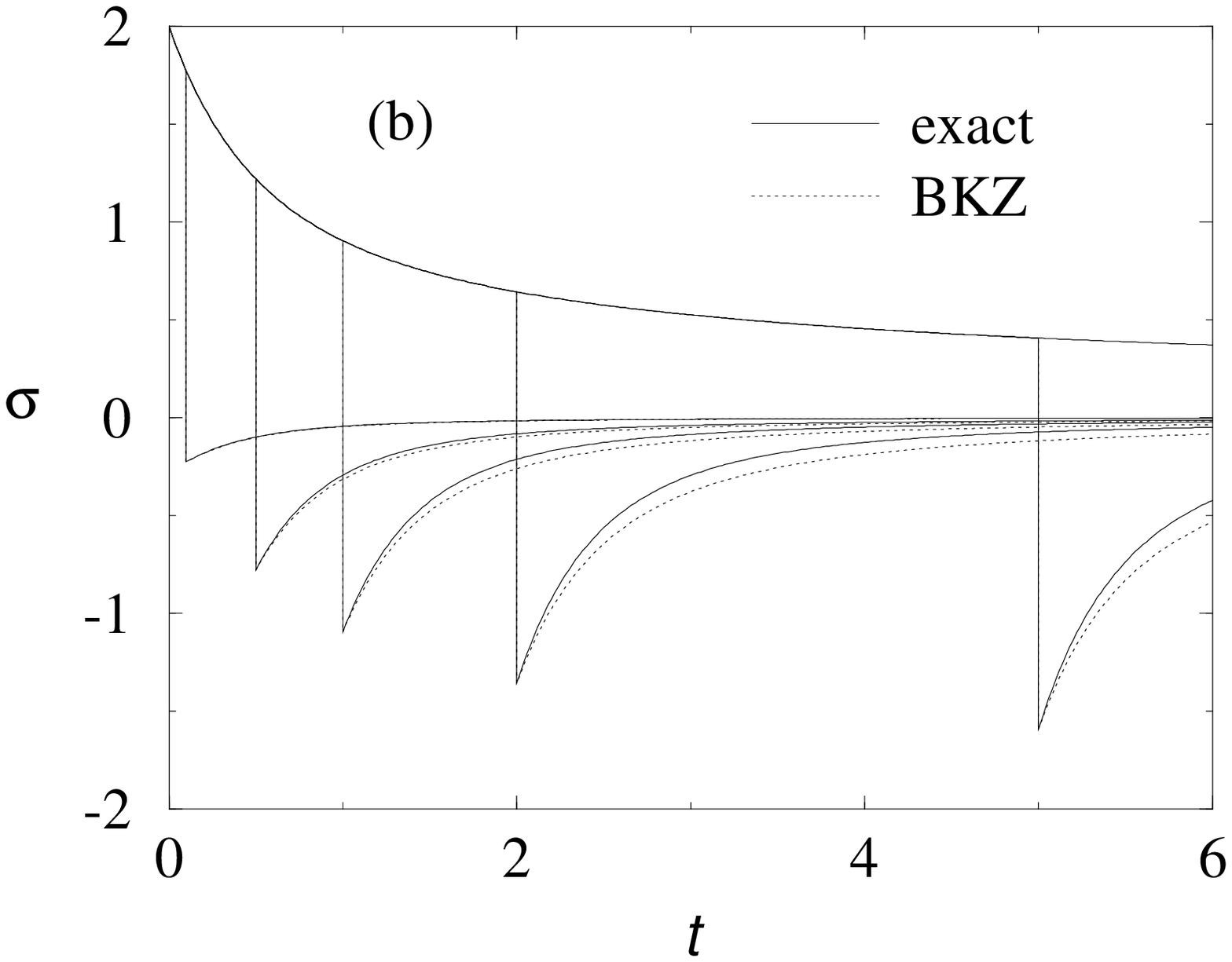}
\efig{two_step_strains}{Stress response to two large step strains of
(a) equal ($\gam_1=\gam_2=2$) and (b) opposite ($\gam_1=-\gam_2=2$) sign,
applied at times $t=0$ and $t=\Delta t=0.1$, 0.5, 1, 2, 5,
respectively. Noise temperature $x=1.5$.}
By applying two (large) step strains in sequence, one can further
probe the nonlinear response of the SGR model. Let $\gam_1$ and
$\gam_2$ be the amplitudes of the two strains. If the first strain is
applied at $t=0$ and the second one at $t=\Delta t$, then
$\gam(t)=\gam_1\Theta(t)+\gam_2\Theta(t-\Delta t)$. It is
straightforward to solve the CE~(\ref{ce},\ref{ce_Gam})
numerically for $t>\Delta t$. Fig.~\ref{fig:two_step_strains}
exemplifies the results for the two cases where the strains are either
equal or of equal magnitude but opposite sign.  In the first case, and
more generally when $\gam_1 \gam_2>0$, the second step strain speeds
up the stress relaxation (by a factor
$\exp\{[(\gam_1+\gam_2)^2-\gam_1^2]/2x\}$ for small $\Delta
t$). Therefore, even though the stress is increased momentarily when
the second strain is applied, it can actually relax back to zero more
quickly than in the absence of this strain.  In the second case
($\gam_1\gam_2<0$), the second step strain can to some degree reverse
the speed-up from the first step strain. A particularly simple form of
the resulting stress response is obtained for $\gam_1=-\gam_2=\gam$
and small $\Delta t$:
\bea
\sig(t>\Delta t) & = &
-\gam \left[1-G\eq\left(e^{\gam^2/2x}\Delta t\right)\right]
\nonumber\\
& & \times\ \Grho\left(e^{\gam^2/2x}(t-\Delta t)\right)
\nonumber
\eea
This can be understood by noting that the stress for $t>\Delta t$ is
due entirely to elements which have yielded between the application of
the first and the second strain; all other elements have simply
followed the two changes of macroscopic strain and are therefore back
to their unstrained state $l=0$ after the second strain. The factor in
squared brackets just gives the fraction of such elements. The time
dependence of the ensuing stress relaxation is determined by $\Grho$
rather than $G\eq$ because elements that have yielded were ``reborn''
with yield energies sampled from $\rho(E)$. These elements---which
have ``forgotten'' about the first step strain---also receive a
speed-up of their relaxation by the second strain.

The above results can be compared to the predictions of the empirical
BKZ (Bernstein, Kearseley, Zapas) equation~\cite{BerKeaZap63}. This
relation approximates the stress response to an arbitrary strain
history in terms of the response $\sig(t)=\phi(t,\gam)$ to a step
strain of size $\gam$ at time $t=0$:
\[
\sig\bkz(t)=\int_{-\infty}^t\!\! dt' \left. 
\deriv{t'}\phi(t-t',\gam) \right|_{\gam=\gam(t)-\gam(t')}
\]
For two step strains, this gives, for $t>\Delta t$
\be
\sig\bkz(t)= \phi(t,\gam_1+\gam_2)-\phi(t,\gam_2)+\phi(t-\Delta t,\gam_2)
\label{bkz_two_steps}
\ee
In our case, $\phi(t,\gam)$ is given by~\(nonlinear_single_step), and
the BKZ prediction is plotted in Fig.~\ref{fig:two_step_strains} along
with the exact results. One finds that for the SGR model, the 
BKZ equation is at best approximate, at worst qualitatively
wrong. This is most easily seen in the size of the stress jump at
$t=\Delta t$; the BKZ equation predicts
\be
\phi(0^+,\gam_2)+[\phi(\Delta t,\gam_1+\gam_2)-\phi(\Delta t,\gam_1)
-\phi(\Delta t,\gam_2)]
\label{BKZ_jump}
\ee 
Because $\phi(0^+,\gam)=\gam$ within the SGR model, the term in square
brackets is the deviation from the true value, which is $\gam_2$. For
$\gam_1=-\gam_2$, the BKZ prediction for the stress jump is exact
because $\phi(t,\gam)=-\phi(t,-\gam)$; in this case
(Fig.~\ref{fig:two_step_strains}b), it also works reasonably well for
the subsequent stress relaxation.  In the general case, however, it is
unreliable; Fig.~\ref{fig:two_step_strains}a shows that it can in fact
even predict the wrong sign for the stress jump.

Finally, we note that a failure of the BKZ equation has also been
observed in double step strain experiments on polymeric
liquids~\cite{DoiEdw86}. There, however, the most pronounced
deviations occur for successive step strains of opposite sign rather
than, as in the SGR model, for strains of the same sign. This can be
understood on the basis of the different kinds of nonlinearities in
the two cases. Roughly speaking, in the polymer case the BKZ equation
fails because it neglects memory of the shape of the tube in which a
given polymer molecule reptates~\cite{DoiEdw86,Doi80}. Such memory
effects are strongest for {\em strain reversal}, which can bring the
tube back to a conformation close to its original shape. In the SGR
model, on the other hand, the BKZ equation fails because it does not
adequately represent the effects of the strain history on the stress
relaxation rates in the material. Such effects are strongest when an
applied strain compounds an earlier speed-up of relaxation processes,
\ie, for double step strains of the {\em same sign}.

\subsection{Large oscillatory strains}
\label{subsec:nonlin_G}

\subsubsection{Dynamic moduli}

As a final example of nonlinear rheological behavior, we consider the
case of large oscillatory strains. We remind the reader at this point
that we have chosen units in which typical local yield strains are of
order unity (see Sec.~\ref{sec:ce}). To transform to experimentally
relevant quantities, all strain values have to be multiplied by a
typical yield strain $\overline{\lc}$ of the SGM under
consideration. A strain $\gam=1$ in our units therefore corresponds to
a real strain of generally at most a few percent.

We consider only the ergodic regime $x>1$; we also ignore transient
behavior caused by start-up of the oscillatory strain. In the steady
state, we can write the stress response to an oscillatory strain
$\gam(t)=\gam\Re e^{i\om t}$ as
\be
\sig(t)=\gam\Re \left[\Gcomp(\om,\gam)e^{i\om t}\right] + \Delta\sig(t)
\label{nonlin_modulus}
\ee
where $\Delta\sig(t)$ contains the contributions from all higher
harmonics. This defines an amplitude dependent dynamic modulus
$\Gcomp(\om,\gam)$; the relative root-mean-square size of the stress
contributions from higher harmonics is measured by the residual $r$,
defined as
\be
r^2=\frac
{\int dt \left[\Delta\sig(t)\right]^2}
{\int dt \ \sig^2(t)}
\label{residual}
\ee
\bfig{9cm}{0cm}{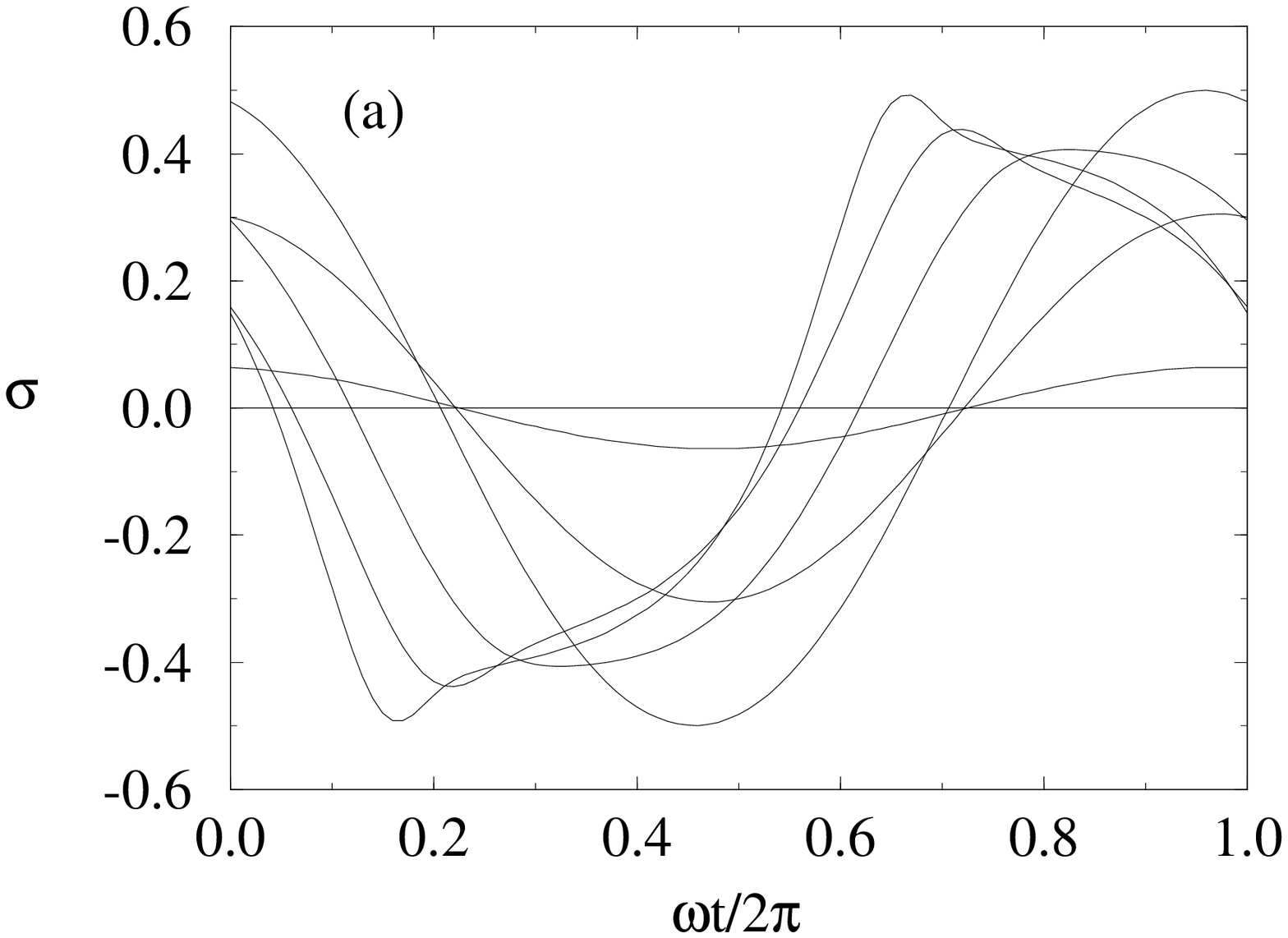}
\end{center}

\vspace*{0cm}
\begin{center}
\epsfxsize 9cm
\epsfbox{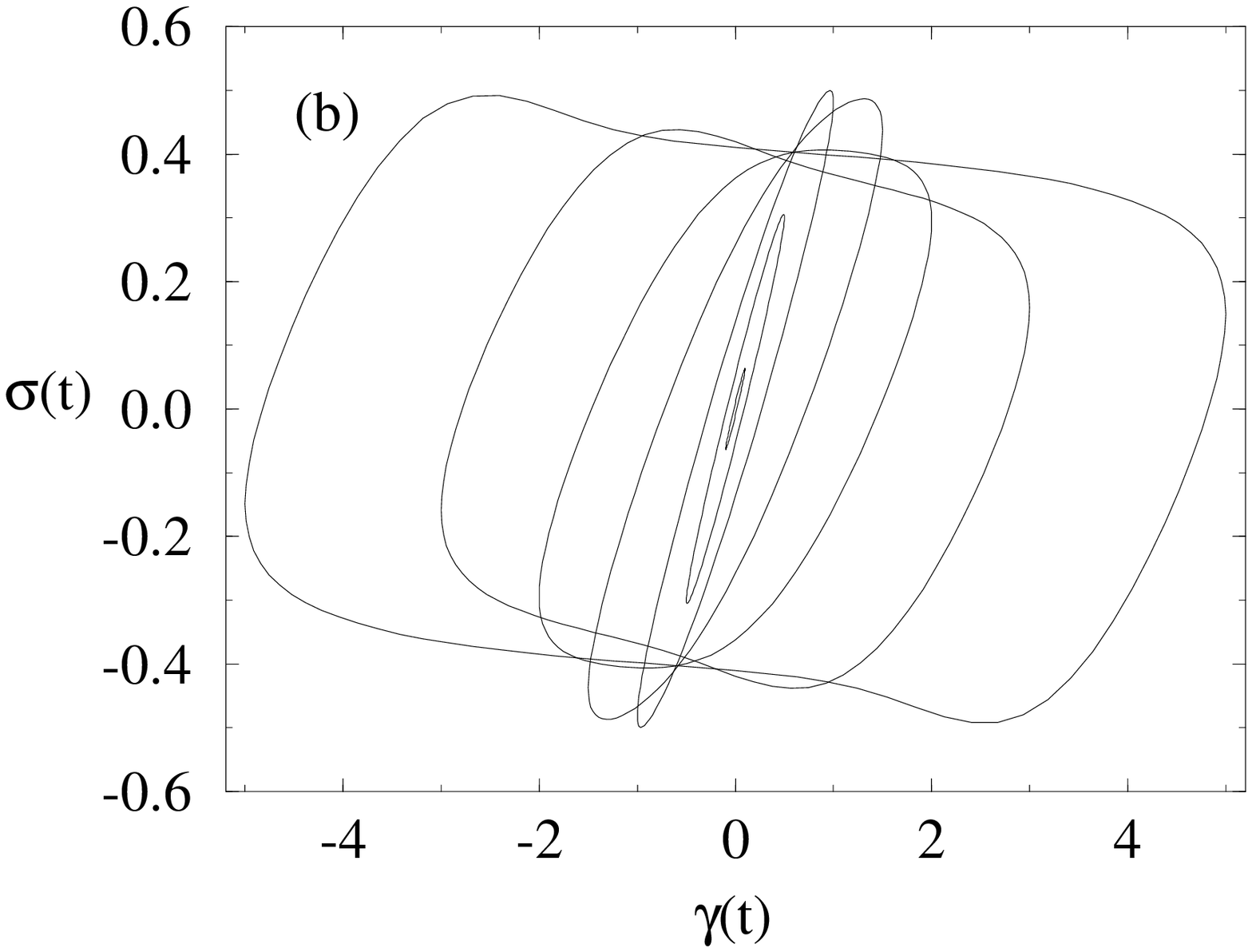}
\efig{sigma_waveform}{(a) Stress response $\sig(t)$ for oscillatory
strain $\gam(t)=\gam\cos(\om t)$, for frequency $\om=0.01$ and
effective temperature $x=1.1$. Initially, the response is almost
perfectly elastic; as the strain amplitude increases (curves are shown
for $\gam=0.1$, 0.5, 1, 2, 3, 5), the zero crossings of $\sig(t)$ move
to the left, corresponding to progressively liquid-like behavior
(strain lagging behind stress). 
(b) Parametric plots of stress $\sig(t)$ vs strain $\gam(t)$, for same
parameter values as in (a); $\gam=1.5$ is also shown.}
\bfig{9cm}{0cm}{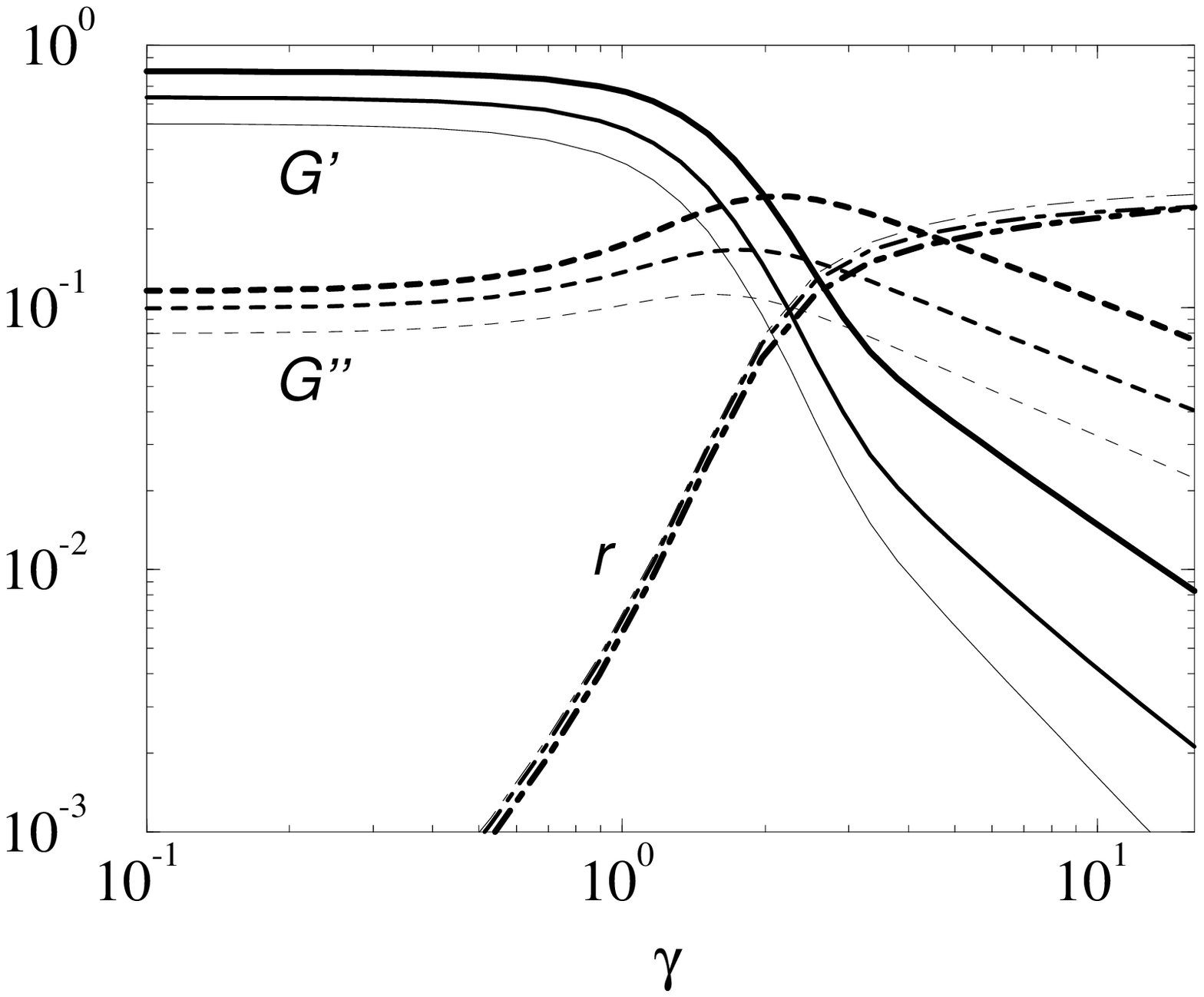}
\efig{strain_sweep}{Strain sweep: Nonlinear moduli $G'$, $G''$ and
residual $r$ as a function of strain amplitude $\gam$. Noise
temperature $x=1.1$; lines of increasing thickness correspond to
$\om=0.001$, 0.01, 0.1. Recall that $\gam$ is rescaled by a typical
local yield strain; $\gam=1$ therefore corresponds to a real strain of
at most a few percent.
}

The determination of $\Gcomp$ and $r$ from the
CE~(\ref{ce},\ref{ce_Gam}) presents no conceptual difficulties, but is
somewhat nontrivial numerically (see App.~\ref{app:nonlin_num} for
details). The solution yields in fact not just $\Gcomp$ and $r$, but
the whole ``waveform'' of the stress response
$\sig(t)$. Fig.~\ref{fig:sigma_waveform}a shows how the response
becomes more and more non-sinusoidal as the strain amplitude is
increased. The stress amplitude first increases linearly with $\gam$,
then drops slightly as the system crosses over from elastic to
liquid-like behavior, and finally rises again slowly as the typical
shear rate $\gam\om$ of the (now essentially liquefied) material
increases. Plotting $\gam(t)$ and $\sig(t)$ in a parametric
stress-strain plot (Fig.~\ref{fig:sigma_waveform}b), one finds a
hysteresis loop for large amplitudes, with stress overshoots near the
points where the strain rate reverses its sign.

Consider now the resulting nonlinear modulus
$\Gcomp$. Fig.~\ref{fig:strain_sweep} shows an example of a ``strain
sweep'': The moduli $G'$ and $G''$ and the residual $r$ are plotted as
a function of strain amplitude for different frequencies $\om$. The
amplitude dependence of $G''$ is particularly noteworthy: As $\gam$
increases, $G''$ first increases, but then passes through a maximum
and subsequently decreases again. This is in qualitative agreement
with recent measurements of nonlinear dynamic moduli in, for example,
dense emulsions and colloidal
glasses~\cite{MasBibWei95,MasWei95,MasLacGreLevBibWei97,BolBibLeq}.
The maximum in $G''$
is most pronounced near the glass transition $x=1$; for higher noise
temperatures, it decreases and disappears altogether around
$x=2$. This is compatible with the following coarse estimate of the
decay of $G''$ beyond the maximum: For sufficiently large strain
amplitudes $\gam$, the system is expected to flow essentially all the
time. If the shear rate $\gamdot$ changes sufficiently slowly ($\om\ll
1$), the stress can be approximated as following ``adiabatically'' the
instantaneous shear rate: $\sig(t) \approx \sig(\gamdot(t))$ with
$\sig(\gamdot)$ the steady shear flow curve. For $1<x<2$ and
sufficiently small shear rates $\gam\omega$, we know from
Sec.~\ref{subsec:steady_shear} that this relationship is a power law,
$\sig(\gamdot)\sim\gamdot^{x-1}$. Hence $\sig(t)\sim(\gam\om\sin\om
t)^{x-1}$ which leads to a $\gam$ dependence of $G''\sim\gam^{x-2}$.
For $x\to 2$, $G''$ should therefore no longer decay for large $\gam$
(as long as the condition $\gam\om\ll 1$ is obeyed), in agreement with
our observation that its maximum with respect to $\gam$ disappears
around this value of $x$. The estimate $G''\sim \gam^{2-x}$ is roughly
compatible with our numerical data, but a precise verification of this
power law is difficult (due to severe numerical problems for $\gam\geq
20$).  Note that within the same approximation, $G'$ would be
estimated to be identically zero, which is of course unphysical.
Instead, we expect it to decay to zero faster than $G''$ as $\gam$
increases, and this is indeed what our numerical data show.

\subsubsection{Size of linear regime}

The above results allow us to determine the size of the linear regime
for oscillatory rheological measurements, \ie, the largest strain
amplitude $\gamc$ for which the measured values of $G'$ and $G''$
represent the linear response of the system.  An important first
observation that can be made on the basis of
Fig.~\ref{fig:strain_sweep} is that the size of the residual $r$ is
not in general sufficient to determine whether one is in the linear
regime or not. For example, for strain amplitude $\gam=1.5$ at $x=1.1$
and $\om=0.1$, $r$ is only around 2.5\% even though the value of $G''$
is already twice as large as in the linear regime. The $\sig(t)$ vs
$\gam(t)$ plot in Fig.~\ref{fig:sigma_waveform}b also demonstrates
this: for $\gam=1.5$, the curve still looks almost perfectly elliptical,
suggesting linear response, while its axis ratio is actually quite
different from the one in the linear regime.  Closer to the glass
transition, this effect becomes even more pronounced. It suggests
strongly that whenever the dynamic moduli of SGMs are measured, an
explicit strain sweep is needed to determine whether measurements are
actually taken in the linear regime.
\bfig{9cm}{0cm}{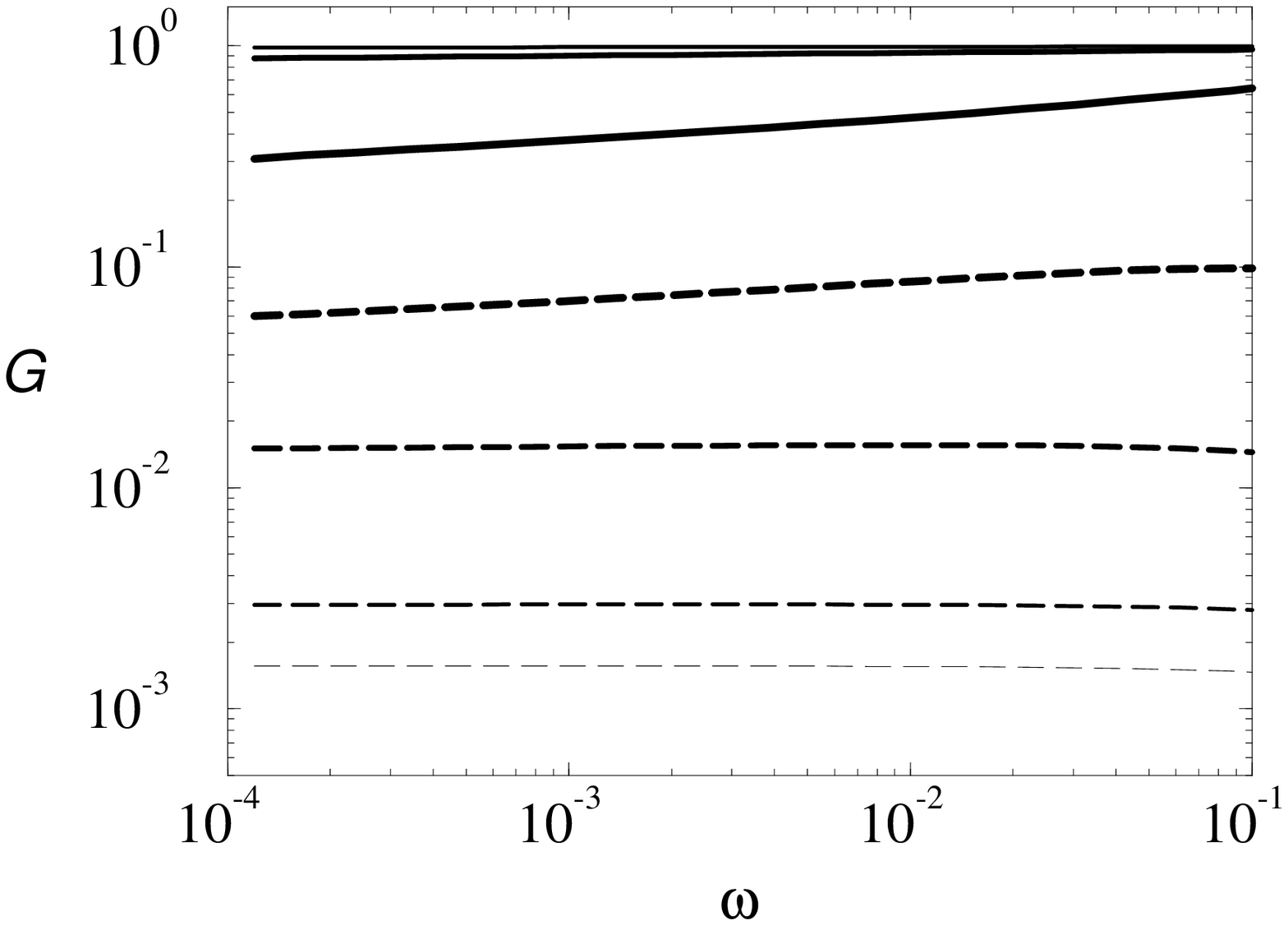}
\efig{const_gam}{Frequency dependence of (nonlinear) dynamic moduli
$G'(\om, \gam)$ (solid lines) and $G''(\om, \gam)$ (dashed) measured
at constant finite strain amplitude $\gam$. Noise temperature
$x=1.001$; increasing values of $\gam=0$, 1, 2, 3 correspond to
increasing line thickness. Recall that $\gam$ is rescaled by a typical
local yield strain; $\gam=1$ therefore corresponds to a real strain of
at most a few percent. The loss modulus $G''$ increases strongly with
$\gam$, whereas $G'$ varies much less (the curves for $\gam=0$ and
$\gam=1$ cannot even be distinguished on the scale of the plot).}

If concerns about nonlinear effects are disregarded, an experimentally
convenient procedure is to measure the dynamic moduli at fixed strain
amplitude $\gam$ (while varying the frequency $\om$). Some numerical
results for this case are shown in Fig.~\ref{fig:const_gam}. Again,
the most interesting behavior occurs near the glass transition. There,
we observe that only relatively minor differences in the amplitude of
the imposed strain can lead to large changes in the measured values of
$G''$ (whereas $G'$ is affected less strongly). This emphasizes again
that extreme caution needs to be taken in experiments designed to
determine the dynamic moduli of soft glassy materials; in particular,
it needs to be born in mind that the loss modulus can easily be
over-estimated due to undetected nonlinear effects.
\bfig{9cm}{0cm}{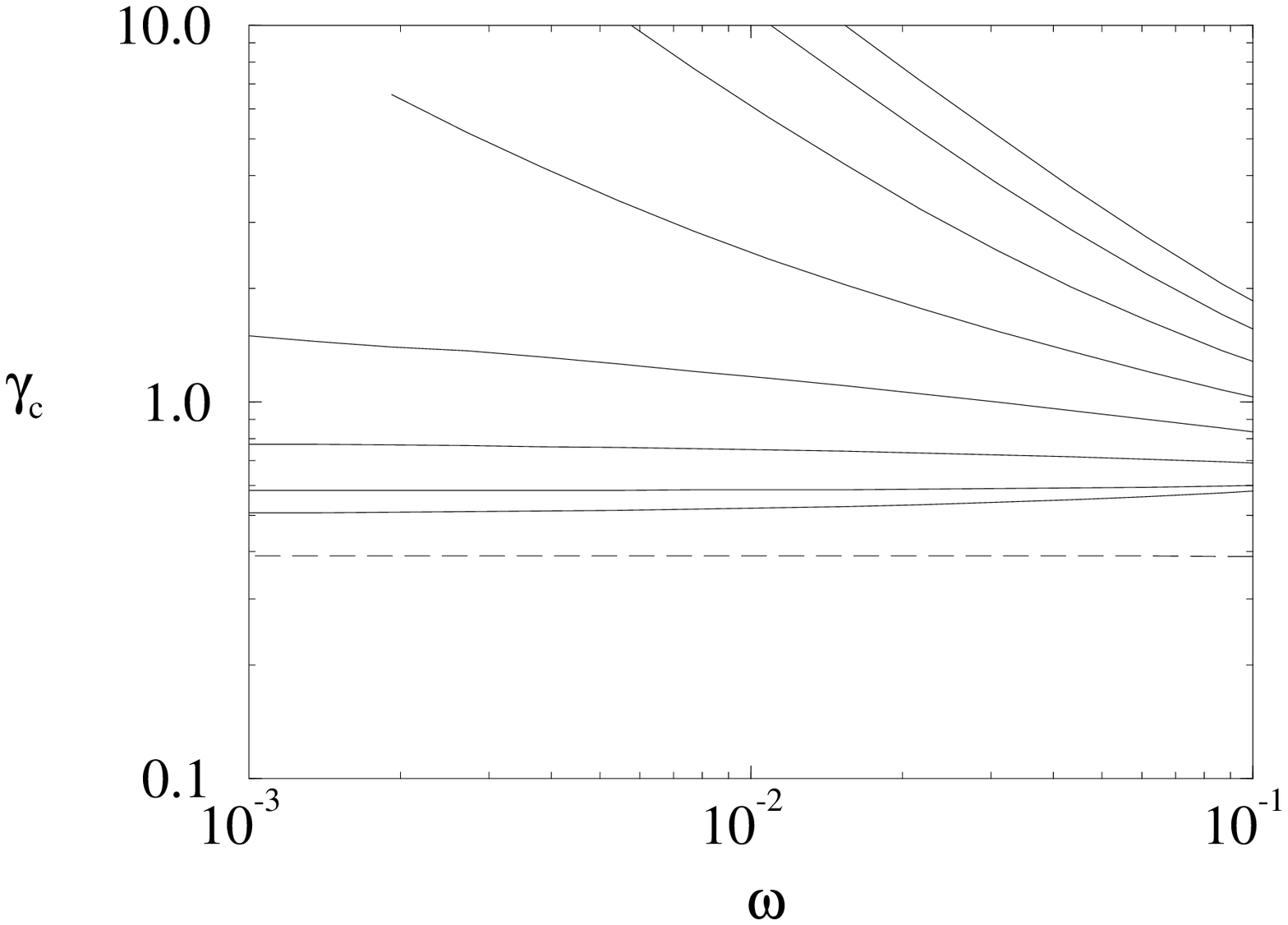}
\efig{gamc}{Size of linear regime $\gamc$ vs
$\om$ for $x=1.001$, 1.5, 2, \ldots, 5 (bottom to top on left). Close
to the glass transition, deviations from linearity first show up in
$G''$, which therefore determines $\gamc$ (dashed line); for larger
$x$, the linear regime is limited by deviations in $G'$ (solid
lines). Recall that $\gamc$, like all strain variables, is rescaled by
a typical local yield strain; $\gamc=1$ therefore corresponds to a
real strain of at most a few percent.
}

Finally, the actual size of the linear regime itself is also of
interest. We choose as a working definition of the linear regime the
strain amplitude $\gamc$ at which either $G'$ or $G''$ first deviate
by 10\% from their values in the limit $\gam\to 0$. (This implies
similar maximum relative deviations for $|\Gcomp|$ and the loss
tangent $\tan\delta=G''/G'$.)  Fig.~\ref{fig:gamc} shows $\gamc(\om)$
for several noise temperatures $x$. Several general trends can clearly
be read off. First, in the low frequency regime, the size of the
linear regime decreases as the glass transition is approached.  This
is intuitively reasonable as one expects nonlinearities to become
stronger near the glass transition~\cite{gamc_wrong_note}. Note,
however, that $\gamc$ does not decrease to zero at the glass
transition; it tends to a finite value of order unity which by our
choice of units corresponds to the typical (a priori) yield stress of
local elements. The frequency dependence of $\gamc(\om)$ also changes
as one moves away from the glass transition: Initially (for $x\approx
1$), $\gamc$ is essentially independent of $\om$ and does remain so
until around $x=3$ (although its absolute value increases); for yet
higher noise temperatures, one finds a crossover to a
$\gamc\sim\om^{-1}$ dependence. The latter corresponds to the
``naive'' criterion that the typical shear rate $\gam\om$ needs to be
smaller than typical relaxation rates (of order unity away from the
glass transition) in order for the imposed strain not to create
nonlinear effects. The predicted $\om$-independence of $\gamc$ near
the glass transition should be easy to verify experimentally.

\section{Interpretation of model parameters}
\label{sec:interpretation}

As has been demonstrated above, the SGR model captures important
rheological features that have been observed in a large number of
experiments, at least in the region around the ``glass transition'' of
the model. Using a mean-field (one element) picture, it is also simple
enough to be generic.  However, a significant challenge that remains
is the interpretation of the model parameters, namely, the ``effective
noise temperature'' $x$ and the ``attempt frequency'' $\Gn$.  To
tackle these questions, we should really start from a more
comprehensive model for the coupled nonlinear dynamics of the
``elements'' of a SGM and then derive the SGR model within some
approximation scheme.  At present, we do not know how to do this, and
the following discussion will therefore have to remain rather
speculative.

\subsection{Effective noise temperature $x$}

We can interpret the activation factor $\exp[-(E-\half kl^2)/x]$ in
the equation of motion~\(basic) of the SGR model as the probability
that (within a given time interval of order $1/\Gn$) a given element
yields due to a ``kick'' from a rearrangement (yield event) elsewhere
in the material. Therefore $x$ is the typical activation energy
available from such kicks. But while kicks can {\em cause}
rearrangements, they also {\em arise} from rearrangements (whose
effects, due to interactions, propagate through the material). So
there is no separate energy scale for kicks: Their energy must of the
order of the energies released in rearrangements, \ie, of the order of
typical yield energies $E$. In our units, this means that $x$ should
be of order unity. Note that this is far bigger than what we would
estimate if $x$ represented true thermal activation. For example, the
activation barrier for the simplest local rearrangement in a foam (a
T1 or neighbor-switching process) is of the order of the surface
energy of a single droplet; this sets our basic scale for the yield
energies $E$. Using typical values for the surface tension and a
droplet radius of the order of $1\mu$m or greater, we find $E\gtappr
10^4 k_{\rm B}T$. In our units $E=O(1)$, so thermal activation would
correspond to extremely small values of $x=k_{\rm B}T\ltappr 10^{-4}$.

We now argue that $x$ may not only be of order one, but in fact close
to one generically. Consider first a steady shear experiment.  The
rheological properties of a sample freshly loaded into a rheometer are
usually not reproducible; they become so only after a period of
shearing to eliminate memory of the loading procedure. In the process
of loading one expects a large degree of disorder to be introduced,
corresponding to a high noise temperature $x\gg 1$. As the sample
approaches the steady state, the flow will (in many cases) tend to
eliminate much of this disorder~\cite{WeaBolHerAre92} so that $x$ will
decrease.  But, as this occurs, the noise-activated processes will
slow down; as $x\to 1$, they may become negligible. Assuming that, in
their absence, the disorder cannot be reduced further, $x$ is then
``pinned'' at a steady-state value at or close to the glass
transition.  This scenario, although extremely speculative, is
strongly reminiscent of the ``marginal dynamics'' seen in some
mean-field spin glass models.  In the spherical $p$-spin glass, for
example, one finds that after a quench from $T=\infty$ to {\em any}
temperature $0<T<T\g$ below the (dynamical) glass transition
temperature $T\g$, the system is dynamically arrested in regions of
phase space characteristic of $T\g$ itself, rather than the true
temperature $T$~\cite{CugKur93,CugKur95}.

There remain several ambiguities within this picture, for example
whether the steady state value of $x$ should depend on $\gamdot$; if
it does so strongly, our results for steady flow curves will of course
be changed.  If a steady flow is stopped and a linear viscoelastic
measurement performed, the results should presumably pertain to the
$x$ characterizing the preceding steady flow (assuming that $x$
reflects structure only).  But unless the strain amplitude is
extremely small the $x$-value obtained in steady state could be
affected by the oscillatory flow itself. This might allow ``flat''
moduli $\Gcomp(\om)$ ($x\approx 1$) to be found alongside a nonzero
yield stress with power law flow exponent around 1/2 ($x\approx
1/2$)~\protect\cite{MasBibWei95,MasBibWei96,PriKis89}.

Experimentally, the above ideas concerning the time evolution of $x$
in steady flows could be tested in systems which can be prepared in
both low- and high-disorder states, such as onion
phases~\cite{onions}: Strain induced ordering starting from an initial
$x$ well below or above $x\g=1$ should drive the system towards $x=0$
or $x\approx 1$, respectively, leading to different rheological
characteristics.

Theoretically, the minimal extension to the SGR model that would be
needed to substantiate the above scenario would be to allow $x$ to
evolve in time. We do not know at present how to deduce the correct
form of this evolution in a principled way from some underlying
microscopic dynamics. However, one possibility is to couple $x$ to the
number of rearrangements in the material, \ie, the yielding rate
$\G$. Indeed, suppose we view $\Gn^{-1}$ as a memory time during which
an element accumulates kicks before attempting a rearrangement. The
number of kicks accumulated is then proportional to $\G/\Gn$. If
individual kicks are thought of as independent Gaussian perturbations,
and we identify $x$ with the mean-squared size of the ``cumulative''
kick, then $x=A\G/\Gn$. The proportionality constant $A$ would depend,
for example, on how kicks propagate through the system. For
$\G/\Gn=1$, each element yields once (on average) within a time
interval $\Gn^{-1}$; $A$ can therefore be viewed as the average number
of kicks caused by a rearrangement. We leave the analysis of such an
approach for future work; preliminary investigations suggest the
emergence of interesting features such as bistable solutions for the
flow curve $\sig(\gamdot)$.

\subsection{Attempt frequency $\Gn$}

Consider now the attempt frequency $\Gn$. It is the only source of a
characteristic timescale in our model (chosen as the time unit above).
This excludes a naive proposal for the origin of $\Gn$: The attempt
frequency cannot be derived (in some self-consistent way) from the
yielding rate $\G$, because the model would then no longer contain an
intrinsic timescale. This would imply that all dependencies on
frequency or time are trivial, leading to unphysical results (the flow
curves $\sig(\gamdot)$ would simply be a constant, as would be the
linear moduli $G'(\om)$ and $G''(\om)$).

We have so far approximated $\Gn$ by a constant value, independently
of the shear rate $\gamdot$; this implies that $\Gn$ is not caused by
the flow directly. One possibility, then, is that $\Gn$ arises in fact
from {\em true} thermal processes, \ie, rearrangements of very
``fragile'' elements with yield energies of order $k_{\rm B}T$.  To a
first approximation, such processes could be accounted for by
extending the basic equation of motion~\(basic) to
\bea
\deriv{t}P(E,l,t) &=&  -
\gamdot\deriv{l}P - \Gloc\, e^{-(E-\half kl^2)/k_{\rm B}T}\,P 
\nonumber\\
& &- \Gn\, e^{-(E-\half kl^2)/x}\,P + \G(t)\,\rh(E)\delta(l)
\label{basic_with_thermal}
\eea
Here $\Gloc$ is an attempt rate for true thermal processes, which
should be a local diffusion rate. In emulsions with $\mu$m droplets,
typical rates for such diffusive modes could be of the order of 1--100
Hz~\cite{BuzLuCat95}. The term on the \rhs\ of~\(basic_with_thermal)
proportional to $\Gloc$ corresponds to yield events caused directly by
thermal fluctuations. Due to the presence of interactions between the
different elements of the material, the effects of such yield events
can propagate through the system and cause other rearrangements. These
are described by the term proportional to $\Gn$.  The ``attempt
frequency'' $\Gn$ is now no longer an independent parameter; instead,
it is proportional to the average rate of thermal rearrangements,
\[
\Gn=A \lav\Gloc\, e^{-(E-\half kl^2)/k_{\rm B}T}\rav_P
\]
The ``propagation factor'' $A$ again represents the number of kicks
caused by a thermally induced yield event. It has a crucial effect on
the behavior of the modified model~\(basic_with_thermal), as can be
seen by considering the equilibrium distribution in the absence of
macroscopic strain ($\gam(t)=0$). One has
$P\eq(E,l)=P\eq(E)\delta(l)$ with
\[
P\eq(E) = \frac{\G}{\Gloc e^{-E/k_{\rm B}T}+\Gn\, e^{-E/x}}\ \rho(E)
\]
When $\Gn$ is of the order of $\Gloc$ or larger,
$P\eq(E)\propto\exp(E/x)\rho(E)$ as in the original version~\(basic)
of the model. From this, the value of $\Gn$ can be calculated; for the
assumption $\Gn\gtappr\Gloc$ to be self-consistent, one then requires
\be
\frac{\Gn}{\Gloc}=A\ \frac{\int \! dE \ \rho(E) \exp(-E/k_{\rm B}T)}
{\int \! dE \ \rho(E) \exp(E/x)} \gtappr 1
\label{A_condition}
\ee
(here we have neglected a term $E/x$ in the exponent of the numerator
because $k_{\rm B}T\ll x$).  This condition can be given an intuitive
interpretation: $A$ must be large enough for each thermal yield event
to produce at least one new element which can yield thermally (\ie,
whose yield energy $E$ is of order $k_{\rm B}T$), thus maintaining the
population of such fragile elements.  For smaller $A$, one finds
instead that $\Gn/\Gloc\sim\exp(-\overline{E}/k_{\rm B}T)$, which for
typical barrier energies $\overline{E}=O(1)$ (in our units) is
unfeasibly slow.  The above mechanism can therefore give a plausible
rheological time scale {\em only} if the average number $A$ of
rearrangements triggered by one local, thermally induced rearrangement
is large enough to sustain the population of fragile elements, as
determined by~\(A_condition).  The values of $A$ actually required for
this are sensitive to the small $E$ behavior of $\rho(E)$. Assuming
for example $\rho(E)\propto E^{y-1}\exp(-E)$, one has the condition
\[
A \gtappr [k_{\rm B}T(1-x^{-1})]^{-y}
\]
For $y=1$, where $\rho(E)$ stays finite for $E\to 0$, this requires at
least $A\gtappr 10^4$. Such large values appear implausible unless a
single yield event could trigger a whole ``avalanche'' of others; in
foams, it has been argued that this might be the
case~\cite{OkuKaw95}. On the other hand, significantly smaller values
of $A$ would be sufficient if $\rho(E)$ shows a significant bias
towards small yield energies $E$ ($0\approx y<1$).  The above
``thermal trigger'' scenario would then be more generically
plausible. To draw more definite conclusions on this point, it would
be useful to measure $\rho(E)$ in, for example, a computer simulation
of a model SGM.

There are a number of other possible explanations for the origin of
$\Gn$.  These include, for example, noise sources internal to the
material, such as coarsening in a foam, or uncontrolled external
noise.  Finally, the rheometer itself could also be a potential
source of noise; this would however suggest at least a weak dependence
of $\Gn$ on the shear rate $\gamdot$. We cannot at present say which
of these possibilities is most likely, nor rule out other
candidates. The origin of $\Gn$ may not even be universal, but could
be system specific.

\section{Conclusion}
\label{sec:conclusion}

We have solved exactly the SGR (soft glassy rheology) model of
Ref.~\cite{SolLeqHebCat97} for the low frequency shear rheology of
materials such as foams, emulsions, pastes, slurries, etc. The model
focuses on the shared features of such soft glassy materials (SGMs),
namely, structural disorder and metastability. These are built into a
generic description of the dynamics of mesoscopic elements, with
interactions represented by a mean-field noise temperature $x$. All
rheological properties can be derived from an exact constitutive
equation.

In the linear response regime, we found that both the storage modulus
$G'$ and the loss modulus $G''$ vary with frequency as $\omega^{x-1}$
for $1<x<2$. Near the glass transition, they become flat, in agreement
with experimental observations on a number of materials.  In the glass
phase, the moduli are predicted to {\em age}; this could provide an
interesting experimental check of the model.

Far above the glass transition, the steady shear behavior is Newtonian
at small shear rates. Closer to the transition ($1<x<2$), we found
power law fluid behavior; in the glass phase, there is an additional
nonzero yield stress (Herschel-Bulkley model). The last two regimes
therefore capture important features of experimental data. Above the
glass transition, the validity of the Cox-Merz rule relating the
frequency dependence of the linear moduli to the shear viscosity can
be checked; it breaks down in the power law fluid region and fails
spectacularly at the glass transition. In this regime, stress
overshoots in shear startup are strongest. We have also calculated the
distribution of energies dissipated in local yield events. At variance
with existing simulation data for foams, this exhibits a shear-rate
dependent crossover between two power-law regimes; this discrepancy
remains to be resolved.

We further probed the nonlinear behavior of the model by considering
large amplitude single and double step strains. The nonlinear response
cannot in general be factorized into strain and time dependent terms,
and is not well represented by the BKZ equation. Finally, we
considered measurements of $G'$ and $G''$ in oscillatory strain of
finite amplitude $\gam$. Near the glass transition, $G''$ exhibits a
maximum as $\gam$ is increased (strain sweep), reproducing qualitative
features of recent measurements on emulsions and colloidal glasses.
The contribution of higher harmonics to the stress response is not
always sufficient to determine whether the response is nonlinear,
emphasizing the need for explicit strain sweeps to get reliable data
in the linear regime. Otherwise, measurements at constant strain
amplitude can lead to strongly enhanced values of the loss modulus
$G''$. Finally, we considered the size of the linear regime itself,
\ie, the largest strain amplitude $\gamc$ at which the measured values
of $G'$ and $G''$ still represent the linear response of the
system. The SGR model predicts that $\gamc$ should be roughly
frequency independent near the glass transition; this point should
also be amenable to experimental verification.

In the final section, we speculated on the physical origin of the most
important parameters of the model, namely, the effective temperature
$x$ and the attempt frequency for rearrangements $\Gn$. We argued that
$x$ should be generically of order unity (in our units). This is
because it represents the typical energy released in a rearrangement,
which is of the same order as the activation energy required to cause
a rearrangement elsewhere in the material. A speculative analogy to
marginal dynamics in other glassy systems suggests that $x$ may in
fact be close to unity in general. This is encouraging, because the
SGR model reproduces the qualitative features of 
experimental data best for $x\approx 1$, \ie,
near the glass transition. We mentioned several hypotheses for the
origin of the attempt frequency $\Gn$, which include events triggered
by thermal fluctuations or internal and external noise sources not
explicitly contained within the model. 

In future work, we plan to explore in more detail the strongly
history-dependent behavior of the model in the glass phase.  Its
simplicity should allow this to be done in detail, thereby providing
the first full theoretical study to be made of the generic
relationship between aging and rheology~\cite{aging_El}. Apart from
this, the main challenge is to incorporate spatial structure and
explicit interactions between elements into the model. This should
help us understand better the mutual dynamical evolution of the
attempt rate, the effective noise temperature and the structural
disorder. In the end, one would hope to derive a model similar to the
present one from such a more microscopic description within some
well-defined approximation scheme.

{\bf Acknowledgements:} The author is indebted to F.~Lequeux,
P.~H\'{e}braud and M.~E.~Cates for their significant contributions to
the development and initial investigation of the SGR model, as
published in~\cite{SolLeqHebCat97}, and for helpful comments on the
present manuscript. Thanks are due also to J.-P.~Bouchaud for several
seminal suggestions. Financial support from the Royal Society of
London through a Dorothy Hodgkin Research Fellowship and from the
National Science Foundation under Grant No.\ PHY94-07194 is gratefully
acknowledged.

\appendix

\section{Derivation of constitutive equation}
\label{app:ce}

The equation of motion~\(basic) of the SGR model can be solved by
making the time-dependent change of variable $l\to\Delta
l=l-\gam(t)$. This eliminates the $\gamdot$ (convective) term,
converting the equation of motion from a PDE to an ODE. Suppressing
the $E$ and $\Delta l$ dependence of $P$, the result reads
\bea
\deriv{t}P(t) &=& - \exp\left\{-\frac{1}{x}\left[E-\half(\Delta
l+\gam(t))^2\right] \right\}\,P(t) 
\nonumber\\
& &+ \G(t)\,\rh(E)\delta(\Delta l+\gam(t))
\nonumber
\eea
This can be integrated to give
\bea
P(t)&=&P(0)\exp\left[-e^{-E/x}z(t,0;\Delta l)\right] \nonumber\\
& &+\int_0^t dt'\, \G(t')\,\rh(E)\,\delta(\Delta l+\gam(t')) \nonumber\\
& &\times \exp\left[-e^{-E/x}z(t,t';\Delta l)\right]
\label{gen_solution}
\eea
with the auxiliary function
\[
z(t,t';\Delta l)=\int_{t'}^t dt'' \exp\left\{
[\Delta l+\gam(t'')]^2/2x\right\}
\]
To simplify matters, we now assume that the initial ($t=0$) state is
completely unstrained, \ie, $\gam(0)=0$ and 
$P(0)=P_0(E)\delta(l)=P_0(E)\delta(\Delta l)$.
The stress can be calculated by
multiplying~\(gen_solution) by $\Delta l$ and integrating over $E$ and
$\Delta l$:
\bea
\sig(t)&=&\gam(t)+\lav\Delta l\rav_{P(t)}
\nonumber\\
       &=&\gam(t) - \int_0^t dt'\, \G(t')\, \gam(t') \int dE\, \rh(E) 
\nonumber\\
       & &\times \exp\left[-e^{-E/x}z(t,t';-\gam(t'))\right]
\label{stress_prelim}
\eea
Here the yielding rate $\G(t)$ is still undetermined, but it can be
got from the condition of conservation of probability: The integral
of~\(gen_solution) over $E$ and $\Delta l$ has to be equal to unity,
hence
\bea
1&=&\int dE\, P_0(E) \exp\left[-e^{-E/x}z(t,0;0)\right]
\nonumber\\
 & &+\int_0^t dt' \,\G(t') \int dE\, \rh(E)
\nonumber\\
 & &\times \exp\left[-e^{-E/x}z(t,t';-\gam(t'))\right]
\label{Gam_prelim}
\eea 
To write the results~(\ref{stress_prelim},\ref{Gam_prelim}) in a more
compact form, the auxiliary functions defined in~\(Gzero_rho_def) and
the abbreviation~\(Z_def)
\bea
\ZZ{t}{t'} &=& z(t,t'; -\gam(t'))
\nonumber\\
&=& \int_{t'}^t dt''
\exp\left\{\left[\gam(t'')-\gam(t')\right]^2/2x\right\}
\nonumber
\eea
are used. This yields directly eq.~\(ce_Gam) for the yielding rate
$\G(t)$, while for the stress one obtains
\be
\sig(t) = \gam(t) -\int_0^t dt'\, \G(t')\,\gam(t')\,\Grho(\ZZ{t}{t'})
\ee
This can be expressed in the more suggestive form~\(ce) by writing
the first term on the \rhs\ as $\gam(t)$
times the \rhs\ of~\(ce_Gam).

\section{Asymptotic behavior of $\Grho(z)$}
\label{app:Grho_asympt}

In this appendix, we derive the asymptotic behavior~\(Grho_asympt) of
$\Grho(z)$. As explained in Sec.~\ref{sec:ce}, our choice of units
$x\g=1$ implies $\rh(\E)=\exp[-\E(1+f(\E))]$ with $f(\E)\to 0$ for
$\E\to\infty$. Hence for any $\delta>0$, there exists $M>0$ such that
$|f(E)|<\delta$ for $E>M$. Our strategy will be to split the defining
integral~\(Gzero_rho_def) for $\Grho(z)$ into two parts, for energies
above and below the threshold $M$ and to bound these
separately. Writing
\bea
\Grho(z) &=& \int_0^M dE\, \rh(E) \exp\left(-ze^{-E/x}\right)
\nonumber\\
& & +\int_M^\infty dE\, \rh(E) \exp\left(-ze^{-E/x}\right)
\nonumber
\eea
the first term on the \rhs\ is trivially bounded by zero from below
and by $\exp[-z\exp(-M/x)]$ from above. The second term, on the other
hand, is bracketed by
\bea
\lefteqn{\int_M^\infty dE\ e^{-(1\pm\delta)E/x}
\exp\left(-ze^{-E/x}\right) = }
\nonumber\\
& & 
x z^{-x(1\pm\delta)} \int_0^{ze^{-M/x}} \!\!dy\ y^{x(1\pm\delta)-1}
e^{-y}
\label{second_bracket}
\eea
(the plus and minus sign giving the lower and upper bound,
respectively).  Now consider the behavior of $\Grho(z)z^{x+\epsilon}$
for some arbitrary small $\epsilon>0$. Choose $\delta=\epsilon/(2x)$
and a corresponding $M$; then from~\(second_bracket)
\[
\Grho(z)z^{x+\epsilon} > x z^{\epsilon/2} \int_0^{ze^{-M/x}} \!\!dy\
y^{x+\epsilon/2-1} e^{-y}
\]
The integral has a finite limit for $z\to\infty$ (it is just a Gamma
function), and so this lower bound tends to infinity in this limit,
proving the first part of~\(Grho_asympt). The second part is
demonstrated in a similar fashion: With the same choice of $\delta$
for a given $\epsilon$, and again using~\(second_bracket),
\bea
\Grho(z)z^{x-\epsilon} &<& z^{x-\epsilon}\exp\left(-ze^{-M/x}\right) 
\nonumber\\
& &+ x z^{-\epsilon/2} \int_0^{ze^{-M/x}} \!\!dy\ y^{x-\epsilon/2-1}
e^{-y}
\nonumber
\eea
Again, the integral has a finite limit (assuming $\epsilon$ is
sufficiently small, \ie, $\epsilon<2x$), and both terms on the \rhs\
therefore tend to zero for $z\to\infty$, completing the proof
of~\(Grho_asympt).

\section{Flow curves and yield stress}
\label{app:yield_stress}

Here we derive the small shear rate behavior of the flow curves
$\sig(\gamdot)$. As shown in Sec.~\ref{subsec:steady_shear}, the
stress $\sig(\gamdot)={I_1(\gamdot)}/{I_0(\gamdot)}$ can be expressed
in terms of the functions
\be
I_n(\gamdot)=\int_0^\infty dl\ l^n\,\Grho(Z(l))
\label{In_def}
\ee
The scaling of $I_n$ with $\gamdot$ can be obtained from the
asymptotic behavior of $\Grho(z)$. From~\(Grho_asympt), it follows
that for any $\epsilon>0$, we can choose a $z_0$ such that
\be
z^{-x-\epsilon}<\Grho(z)<z^{-x+\epsilon} \quad \mbox{for $z>z_0$}
\label{Grho_bound}
\ee
Now we use $z_0$ to decompose the $l$-integral in~\(In_def) into the
parts with $l\lessgtr z_0\gamdot$:
\[
I_n = I_n^< + I_n^> \qquad I_n^< = \int_0^{z_0\gamdot} dl\ l^n\,\Grho(Z(l))
\]
Replacing $\Grho(Z(l))$ by its minimum and maximum over the
integration range, $I_n^<$ is trivially bounded by
\[
\Grho(Z(z_0\gamdot))<\frac{n+1}{(z_0\gamdot)^{n+1}} I_n^< < 1
\]
As $\gamdot\to 0$, the \lhs\ tends to $\Grho(z_0)$, so we have the
scaling $I_n^< = O(\gamdot^{n+1})$. To bound $I_n^>$, we use that
$Z(l)>l/\gamdot>z_0$ in the relevant integration range, so that the
bounds~\(Grho_bound) on $\Grho$ can be used. Writing $Z(l)$ out
explicitly, this gives lower and upper bounds for $I_n^>$ of
\[
\gamdot^{x\mp\epsilon} \int_{z_0\gamdot}^\infty dl\ l^n\, 
\left(\int_0^l d\gam\, e^{\gam^2/2x}\right)^{-x\pm\epsilon}
\]
For $x<n+1$ (and $\epsilon$ sufficiently small), the outer integral
has a finite limit for $\gamdot\to 0$, and so $I_n^>$ scales as
$\gamdot^x$ up to sub-power law factors. For larger values of $x$, on
the other hand, this integral diverges as
$\gamdot^{n+1-x\pm\epsilon}$.  $I_n^>$ then scales as $\gamdot^{n+1}$
(since both the lower and upper bound do), \ie, in the same way as
$I_n^<$.

As discussed in Sec.~\ref{subsec:steady_shear}, the above scaling
properties of $I_n^<$ and $I_n^>$ prove that the flow curve is a power law
$\sig\sim\gamdot^{x-1}$ (up to sub-power law factors) in the regime
$1<x<2$. In the glass phase ($x<1$), the simplest case is that of
exponential $\rho(E)$ (eq.~\(exp_rho)). The asymptotic behavior of
$\Grho(z)\sim z^{-x}$ then translates directly into
$I_n^>\sim\gamdot^x$ without sub-power law corrections, and this gives the
Herschel-Bulkley form~\(HB_exp_rho) of the flow curve. The yield
stress~\(sigy_exp_rho) is given by the limit of $I_1^>/I_0^>$ for
$\gamdot\to 0$, while the power law onset of the additional stress
arises from the small corrections due to $I_0^<$.

For general $\rho(E)$, on the other hand, the sub-power law factors in
$I_n^>(\gamdot)$ cause a corresponding weak $\gamdot$ dependence of
$\sig(\gamdot)$, which dominates the effect of the small correction
terms $I_n^<(\gamdot)$. The flow curve therefore no longer has
the simple Herschel-Bulkley form~\(HB_exp_rho). However, in the
examples that we tested numerically ($\rho(E)\sim E^n \exp(-E)$ for
$n=1$, 2, 3), we found that this form still provides a good fit to
$\sig(\gamdot)$ over several decades of shear rate $\gamdot$. Both the
exponent and yield stress of such a fit are then only effective
quantities and depend on the range of $\gamdot$ considered; they are
therefore no longer directly related to $x$. In the examples that we
studied, we always found values of the effective exponent
significantly below unity.

The slow sub-power law variation of $\sig(\gamdot)$ for general
$\rho(E)$ means that there is, for practical purposes, always an
effective yield stress (whose actual value depends weakly on the
lowest accessible shear rate $\gamdot$). Nevertheless, one may wonder
what the ``true'' yield stress $\sigy=\sig(\gamdot\to 0)$ would be.
The above line of argument does not answer this question; it does not
even exclude the possibility of $\sigy$ being zero. We have examined
this issue for several different sub-power law corrections to the
asymptotic behavior of $\Grho$, such as $\Grho(z)z^x\sim(\ln z)^m$, or
$\sim \exp[(\ln z)^n]$ with $|n|<1$. The yield stress is always
nonzero, and in fact turns out to be the same as for exponential
$\rho(E)$. We suspect that this may be true in general, but have not
found a proof.

\section{Numerical determination of $\Gcomp(\om,\gam)$}
\label{app:nonlin_num}

In this appendix, we outline the numerical scheme that we used to
obtain the nonlinear dynamic modulus $\Gcomp(\om,\gam)$ and the
residual $r$ defined in~\(nonlin_modulus) and~\(residual),
respectively. As explained in Sec.~\ref{subsec:nonlin_G}, we are
interested in the steady state stress response in the ergodic regime
$x>1$. We can then safely send the initial time to $-\infty$ in the
CE~(\ref{ce},\ref{ce_Gam}). The equations that need
to be solved can be simplified further by using the fact that in the
steady state, the yielding rate $\G(t)$ must have the same periodicity
as the applied strain $\gam(t)$. Denoting the oscillation period by
$T=2\pi/\om$, the task is then to solve
\be
1=\int_{t-T}^t dt'\ \G(t') H(t,t')
\label{ce_Gam_nonlin}
\ee
for $\G(t)$ and then to evaluate the stress from
\be
\sig(t)=\gam(t)-\int_{t-T}^t dt'\ \gam(t') \G(t') H(t,t')
\label{ce_nonlin}
\ee
Here the periodicity of the problem has been absorbed into the
definition of 
\bea
H(t,t') &=& \sum_{n=0}^\infty \Grho(\ZZ{t}{t'-nT})
\nonumber\\
&=& \sum_{n=0}^\infty \Grho(\ZZ{t}{t'}+n\ZZ{t'+T}{t'})
\nonumber
\eea
where the second equality follows again from the periodicity of the
strain $\gam(t)=\gam\cos\om t$. The numerical solution of the integral
equation~\(ce_Gam_nonlin) is simplified by subtracting from the kernel
$H(t,t')$ a part that depends on $t'$ only:
\[
\tilde{H}(t,t')=H(t,t')-H(t'+T,t')=\lav\frac{e^{-\Omega
Z_1}-e^{-\Omega Z_2}}{1-e^{-\Omega Z_2}}\rav_\rho
\]
where we have abbreviated $\Omega=\exp(-E/x)$, $Z_1=\ZZ{t}{t'}$, $Z_2 =
\ZZ{t'+T}{t'}$. The modified kernel $\tilde{H}(t,t')$ has the
convenient properties $\tilde{H}(t',t')=1$, $\tilde{H}(t'+T,t')=0$ and
is also simpler to evaluate numerically than $H(t,t')$.  The yielding
rate can easily be calculated from $\tilde H$ instead of $H$: Defining
a modified yielding rate $\tilde{\G}(t)$ as the solution of
\be
1=\int_{t-T}^t dt'\ \tilde{\G}(t') \tilde{H}(t,t')
\label{ce_Gam_mod}
\ee
the actual yielding rate is recovered by dividing by the constant
factor
\[
1+\int_{0}^T dt'\ \tilde{\G}(t') H(t'+T,t')
\]
However, even the solution of~\(ce_Gam_mod) is still nontrivial,
especially in the low frequency regime $T\gg 1$ that we are most
interested in. This is because $\tilde{H}$ ``inherits'' from $\Grho$ an
initial ``fast'' decay as $t-t'$ increases from zero, followed by a much
slower power-law decay (which in turns gives way to a rapid final
decay as soon as strain-induced yielding becomes important). This
separation of $O(1)$ and $O(T)$ timescales rules out traditional
solution methods such as Chebyshev approximation. Instead, we
solve~\(ce_Gam_mod) by Fourier transform: Writing
\[
\tilde\G(t)=\sum_{n=-\infty}^\infty \tilde\G_n e^{in\om t}
\]
eq.~\(ce_Gam_mod) is transformed into the matrix equation
\be
\sum_{m=-\infty}^\infty \tilde H_{nm}\tilde\G_m=\delta_{n,0}
\label{ce_Gam_Fourier}
\ee
with coefficients
\[
\tilde H_{mn}=\int_0^T \frac{dt}{T}\ e^{-i(n-m)\om t}\int_0^T d\tau\
e^{-im\om\tau} \tilde H(t,t-\tau)
\]
Once~\(ce_Gam_Fourier) is solved and the rescaling from
$\tilde\G$ to $\G$ is carried out, the stress is obtained as
\bea
\frac{\sig(t)}{\gam} &=& \sum_n \sig_n e^{in\om t}
\nonumber\\
\sig_n &=& \half(\delta_{n,-1}+\delta_{n,1}) - \half \sum_{m} \G_m 
(\tilde H_{n,m+1} + \tilde H_{n,m-1})
\nonumber
\eea
Its Fourier components determine the nonlinear dynamic modulus and
squared residual as 
\[
\Gcomp(\om,\gam)=2\sig_1 \qquad
r^2=1-\frac{|\sig_1|^2}{\sum_{k=0}^\infty |\sig_{2k+1}|^2}
\]
The result for $r^2$ has been simplified using the fact that
$\sig_{-n}=\sig^*_n$ (because $\sig(t)$ is real) and that $\sig_n=0$
for even $n$ (because $\sig(t)\to-\sig(t)$ for $\gam\to-\gam$, which
corresponds to $t\to t+T/2$).

To solve the main equation~\(ce_Gam_Fourier), we truncate the matrix
equation at successively higher orders until the calculated values of
$G'(\om,\gam)$, $G''(\om,\gam)$ and $r$ are stable to within 1\%. The
Fourier components $\tilde H_{mn}$ are calculated from a spline
interpolant approximation to $\tilde H(t,t')$ in order to save
expensive function evaluations.


\begin{thebibliography}{10}
\bibitem{emulsion_inherent_metastability}
Soft systems may also be intrinsically metastable in a more drastic sense (for
  example, with respect to coalescence in emulsions)---we ignore this here.
\bibitem{strong_fragile_note}
(Structural) glasses are often classified into ``strong'' and ``fragile'' glass
  formers (see \eg~\protect\cite{Angell95}). The former can be characterized,
  for example, by an Arrhenius-dependence of the viscosity $\eta$ on $T$
  temperature, while the latter show a much more dramatic variation of $\eta$
  with $T$ near the glass transition. Mode-coupling theories generally predict
  fragile behavior, while trap models have been associated with strong glasses
  (see \eg~\protect\cite{vanMegUnd94}). The latter is only true for
  approximately uniform trap depths, however; indeed, we shall find within the
  SGR model a strongly non-Arrhenius dependence of $\eta$ on the (noise)
  temperature $x$, due to a large spread of trap depths.
\bibitem{Epositive_footnote}
It is always understood in the following that trap depths are restricted to
  positive values, $E>0$.
\bibitem{sawtooth_note}
Precisely this motion is predicted, on a global rather than mesoscopic scale,
  for perfectly ordered foams (see \eg~\protect\cite{Kraynik88}).
\bibitem{beta_relax_footnote}
Within mode-coupling theory, the fact that $\beta$-relaxations correspond to
  localized motion---loosely describable as the rattling of particles in the
  cages formed by their neighors~\protect\cite[p.~333]{GoetSjoe92}---has been
  deduced from the fact that correlation functions factorize into a time
  dependent and a separation (or wavevector) dependent part. Nevertheless, some
  degree of cooperative motion must be involved, at least on a local scale,
  because the motion of each particle within its cage affects the cages of its
  neighbors. The SGR model only captures the analog of the slower
  $\alpha$-relaxations, corresponding to the breakup of particle cages and
  structural rearrangements.
\bibitem{no_diss_in_eq}
This property of the model can be seen most clearly in the equilibrium state
  ($\gamdot=0$) for $x>1$. All elements are then unstrained, so both terms on
  the \rhs\ of~(\protect\ref{energy_balance}) are zero: The total elastic
  energy is constant (and equal to zero) and no energy is being dissipated. At
  the same time, however, rearrangements are predicted to occur at a finite
  rate $\G$.
\bibitem{rho_E_footnote}
This is true for a fixed distribution of yield energies $\rho(E)$, such as the
  one that we use in all our numerical work ($\rho(E)=\exp(-E)$). In general,
  our choice of $x\g=1$ fixes only the asymptotic behavior
  $\rho(E)\sim\exp(-E)$, but not the details of the distribution of small yield
  energies $E$. These do affect the quantitative predictions of the SGR model,
  but do not alter qualitative features such as the power-law behavior of the
  shear moduli and flow curves.
\bibitem{lower_limit_zero_note}
The lower limit in the response integral~(\protect\ref{lin_response}) is zero
  rather than $-\infty$ because we assumed that the system is completely
  unstrained at $t=0$.
\bibitem{num_error_footnote}
The corresponding curves in Ref.~\cite{SolLeqHebCat97} (Fig.~2) contained a
  numerical error. However, this only caused slight changes in the quantitative
  details of the flow curves, without altering any of the qualitative
  conclusions (power laws etc).
\bibitem{P_de_bump_note}
For $\gamdot\gtappr\sqrt{x}$, $P(\de)$ can actually be non-monotonic, with a
  maximum for values of $\de$ of order unity. This is due to the yielding of
  elements from ``metastable'' states with $E-\half l^2<0$.
\bibitem{shear_startup_Emax_note}
The behavior in the glass phase ($x<1$) could be investigated by introducing an
  energy cutoff $\Emax$ as in Sec.~\protect\ref{subsec:linear_glass}. However,
  the transient behavior $\sig(t)$ turns out to be very sensitive to $\Emax$
  (whereas the steady state stress is not) and is therefore of questionable
  significance.
\end{thebibliography}

\end{document}